\documentclass[a4paper,11pt]{article}
\usepackage{jheppub} % for details on the use of the package, please see the JINST-author-manual
\usepackage{lineno}
\usepackage{bbm}
\usepackage{amsmath}
\usepackage{amssymb}
\usepackage{braket}
\usepackage{bm}
\usepackage{xcolor}
\usepackage{ dsfont }
\usepackage{graphicx}
\usepackage{physics}
\usepackage{tikz}
\usetikzlibrary{arrows.meta}
\usetikzlibrary{positioning}
\usetikzlibrary{shapes.misc}
\usetikzlibrary{decorations.markings}
\usetikzlibrary{calc}
\tikzset{cross/.style={cross out, draw=black, ultra thick, minimum size=3*(#1-\pgflinewidth), inner sep=0pt, outer sep=0pt},
cross/.default={5pt}}

\title{\boldmath Monodromy defects in ABJM theory and integrability}

\author{Charlotte Kristjansen,}
\author{and Andrey Shusharin}
\affiliation{Niels Bohr International Academy, Niels Bohr Institute, Copenhagen University, \\
Blegdamsvej 17, DK-2100 Copenhagen \O, Denmark}

\emailAdd{kristjan@nbi.dk}
\emailAdd{andrey.shusharin@nbi.ku.dk}

\abstract{ 
We establish that supersymmetric monodromy defects in ABJM theory, also known as vortex loops, define integrable boundary states of the underlying alternating $SU(4)$ spin chain. Exploiting the KT relation within the framework of the algebraic Bethe ansatz, we derive a closed-form expression for the leading-order one-point functions of non-protected operators in the presence of these defects. As a special case, our result reproduces the spin-chain overlap governing the three-point functions of two maximal giant gravitons and a single tiny graviton. We furthermore investigate the quantization of the theory in the defect background and compute the first quantum correction to the one-point functions in the simplest setting. Finally, we discuss possible extensions of our overlap formula.
}

\begin{document}
\maketitle
\flushbottom

\section{Introduction}

Monodromy defects in the AdS/CFT correspondence constitute a particularly intriguing class of defects. In contrast to conventional defects, they introduce non-trivial field monodromies, making them both conceptually and technically more challenging, and consequently less explored.

A famous example is provided by the 1/2--BPS Gukov-Witten~\cite{Gukov:2008sn,Gukov:2006jk}
surface defects in ${\cal N}=4$ Super Yang-Mills theory.
Here the gauge field explicitly exhibits a holonomy around the defect. Gukov-Witten surface defects were originally studied for their relevance for the geometric Langlands program and for the understanding of S-duality. They were shown to admit interesting string theory realizations~\cite{Drukker:2008wr} and  later became central ingredients in supersymmetric localization and the AGT correspondence~\cite{Gukov:2014gja}. More recently,
 localization techniques have been applied directly to the original Gukov-Witten set-up in order to compute expectation values of protected operators and supersymmetric Wilson loops in the defect background~\cite{Choi:2024ktc}. In a parallel line, Gukov-Witten defects were studied by integrability techniques in~\cite{Holguin:2025bfe,Chalabi:2025nbg} which made it possible to address expectation values of non-protected operators. Finally, the application of yet
 another exact method to the Gukov-Witten defects, that of analytical bootstrap, is promising
 but has so far only been explored to a limited extent~\cite{Holguin:2025bfe,BianchiTalk}.

Alongside these recent developments in the study of Gukov-Witten surface defects, the substantial progress made in the treatment of conventional defects, see e.g.~\cite{Andrei:2018die}, has led to an enhanced interest in monodromy defects. Monodromy defects can be viewed as higher-dimensional generalizations of the twist fields familiar from two-dimensional conformal 
field theories~\cite{Dixon:1985jw,Dixon:1987qvj}. Monodromy defects in the $d$-dimensional $O(N)$ model and in free theories were recently addressed in respectively e.g.~\cite{Soderberg:2017oaa,Giombi:2021uae} and~\cite{Lauria:2020emq,Bianchi:2021snj}. Furthermore,  a novel type of monodromy defects 
in ${\cal N}=4$ SYM, denoted as charge-conjugation defects were introduced in~\cite{Gomis:2025gzb}.

We shall discuss a lower-dimensional analogue of the (non-rigid) Gukov-Witten surface defects, namely co-dimension two monodromy defects in ABJM theory. These defects were introduced in~\cite{Drukker:2008jm} where they were denoted as vortex loops, the name originating from the fact that some fields exhibit vortex like singularities in the vicinity of the one-dimensional defect. 
They come in different versions with varying amounts of supersymmetry, being 1/2--BPS, 1/3--BPS or
1/6--BPS configurations. 
More general monodromy defects in Chern-Simons theories have been studied in~\cite{Ambrosino:2026ovo}.

We shall design boundary states that represent the supersymmetric monodromy defects 
in the underlying spin chain language of ABJM theory~\cite{Minahan:2008hf}, and we shall demonstrate that these constitute integrable boundary states. Furthermore, by making use of the so-called
KT-relation derived in the algebraic Bethe ansatz approach to integrability~\cite{DeLeeuw:2019ohp,Gombor:2021hmj,Gombor:2024iix,Gombor:2025wvu}
we shall carry out a completely analytic derivation, which includes a certain regularization procedure, of the spin-chain overlap needed to determine the
one-point functions of non-protected operators in the defect background
at  leading loop order.
As a consequence, a closed form expression encompassing the
leading loop order contribution to all such one-point functions  will be presented. 

Other integrable defects in ABJM theory have previously been identified. A class of domain wall defects was shown to be integrable in~\cite{Gombor:2021hmj,Kristjansen:2021abc,Gombor:2022aqj}, while~\cite{Yang:2021hrl} identified an integrable boundary state corresponding to the insertion of a giant graviton. More recently, 
~\cite{Jiang:2023cdm} showed that a subclass of supersymmetric Wilson loops likewise gives rise to integrable boundary states. In all these cases, the one-point functions could be obtained in closed form. Our result encompasses the giant graviton case, as well as certain Wilson loops, as special cases.
Searches
for integrable boundary states of the integrable alternating $SU(4)$ spin chain without reference to defects or AdS/CFT was carried out in~\cite{Bai:2024qtg,Liu:2025uiu,Liu:2026fxi,Bai:2026hun}.

Our paper is organized as follows. We begin in section~\ref{monodromydefects} by introducing the supersymmetric monodromy defects of ABJM theory following the original paper~\cite{Drukker:2008jm}. Subsequently, in section~\ref{onepoint}
we present the matrix product states representing these defects
in the spin chain language. We also briefly describe the
tools of integrability needed to understand the  KT-relation and afterwards proceed
to the solution of this equation. At the end of the section we present the final closed form expression for the leading order one-point functions of non-protected operators in the monodromy backgrounds 
and demonstrate that it contains as a special case the result for the three-point function of two maximal giant gravitons
and a single tiny graviton as well as the result for certain Wilson loops.
The ensuing section~\ref{quantization} contains a discussion of the
strategy and the challenges of quantizing the theory in the monodromy background. The quantization is carried through for the simplest set-up and the leading quantum correction to
one-point functions in this set-up is obtained. In Section~\ref{extension} we discuss a possible extension of our overlap formula, and finally 
Section~\ref{conclusion} contains
our conclusion. Certain details are relegated to appendices.

\label{sec:intro}

\section{The supersymmetric monodromy defects \label{monodromydefects}}

What we will refer to as monodromy defects in ABJM theory were constructed in~\cite{Drukker:2008jm} where they were denoted as vortex loops.
These objects constitute line defects in spacetime along which some of the scalar fields exhibit vortex-like singular profiles, and an accompanying gauge connection 
produces a non-trivial holonomy around the defect. Furthermore, requiring BPS nature as well as scale invariance of the vortex configuration implies that the scalar fields acquire a monodromy around the
defect. 

Introducing the defects we will follow the original paper~\cite{Drukker:2008jm}. 
The point of departure is ABJM theory, a Chern-Simons-matter theory with gauge group $U(N)_k\times \hat{U}(N)_{-k}$, where the subscripts denote the Chern-Simons levels. For $k\neq 1,2$, ABJM theory has ${\cal N}=6$ superconformal symmetry, with 24 supercharges being the fermionic generators of the $OSp(4|6)$ superconformal algebra, whose bosonic subalgebra is $SO(2,3)\times SU(4)$, corresponding to conformal and R-symmetry transformations, respectively.
For our analysis we will only need the part of its action that involves  bosonic fields alone
which reads~\cite{Minahan:2009te,Bandres_2008} 
\begin{equation}\label{ABJM action}
\begin{aligned}
\mathcal{L}_{bos}=  \frac{k}{4 \pi} \operatorname{tr}&\left[\varepsilon^{\mu \nu \lambda}\left(-A_\mu \partial_\nu A_\lambda-\frac{2i}{3} A_\mu A_\nu A_\lambda+\hat{A}_\mu \partial_\nu \hat{A}_\lambda+\frac{2i}{3} \hat{A}_\mu \hat{A}_\nu \hat{A}_\lambda\right)\right.  +D_\mu Y_A^{\dagger} D^\mu Y^A
 \\
&+\frac{1}{12} Y^A Y_A^{\dagger} Y^B Y_B^{\dagger} Y^C Y_C^{\dagger}+\frac{1}{12} Y^A Y_B^{\dagger} Y^B Y_C^{\dagger} Y^C Y_A^{\dagger} \\
& -\frac{1}{2} Y^A Y_A^{\dagger} Y^B Y_C^{\dagger} Y^C Y_B^{\dagger}+\frac{1}{3} Y^A Y_B^{\dagger} Y^C Y_A^{\dagger} Y^B \left.Y_C^{\dagger}\right],\\
\end{aligned}
\end{equation}
where $A=1,2,3,4$ are $SU(4)$ R-symmetry indices.  The $Y$-fields are complex scalars which transform in the bi-fundamental of the gauge group whereas $A$ and $\hat{A}$ are gauge fields which transform in the adjoint representation of respectively
$U(N)_{k}$ and $\hat{U}(N)_{-k}$.  The covariant derivative is defined as 
\begin{equation} \label{adjoint}
    D_{\mu}X=\partial_{\mu}X + i A_{\mu}X -iX\hat{A}_{\mu},\ D_{\mu}X^{\dagger}=\partial_{\mu}X^{\dagger} + i \hat{A}_{\mu}X^{\dagger}- iX^{\dagger}A_{\mu}.
\end{equation}
The theory has a planar 't Hooft limit with $k,N\rightarrow \infty, \lambda=\frac{N}{k}$ fixed. The  monodromy defects are defined  without any reference to this limit.

 A vortex loop is a particular example of a disorder operator. It is a curve in spacetime along  which certain fields become singular. The expectation value of the disorder operator is defined by a path integral over fields obeying the prescribed singular boundary conditions along the defect. Perturbatively, this amounts to quantizing the theory around the corresponding singular classical field configuration.
 
The vortex loops introduced in~\cite{Drukker:2008jm}  are found by requiring three conditions to be fulfilled. The singular field configuration should fulfil the theory's equations of motion, it should conserve some fraction of the supercharges of the parent theory, and it should be scale invariant. For the derivation it proves convenient to introduce 
\begin{equation}
A^{\pm}_\mu= A_\mu\pm \hat{A}_\mu.
\end{equation}
One finds three classes of monodromy defects which are respectively 1/2 BPS, 1/3 BPS and 1/6 BPS. The solutions are most easily explained in the case with $U(1) \times U(1)$ gauge symmetry. In addition, it
is convenient to place the defect along the $t$-axis and work with complex transverse coordinates $z,\bar{z}$. 
We note that a straight line  preserves an $SO(2,1)\times SO(2)\simeq SU(1,1)\times U(1)_L$  subgroup of the three-dimensional conformal group, corresponding to conformal transformations along the line and rotations in the transverse plane. With this set-up the three solutions (denoted by calligraphic letters) read as follows.

\paragraph{1/2 BPS:} In this case one of the four of the complex scalars plus the gauge field are non-vanishing, more precisely
\begin{equation} \label{1/2BPS}
\mathcal{Y}^1=\frac{\beta}{\sqrt{z}}, \hspace{0.5cm} \mathcal{A}_z^+ =- i \frac{\alpha}{2 k z}, \hspace{0.5cm} \mathcal{A}_t^+= -\frac{\beta^2}{|z|}. 
\end{equation}
Here $\alpha$ and $\beta$ are free parameters which can both be taken to be real (by exploiting the gauge invariance of
the system). Furthermore, to ensure invariance under large gauge transformations the parameter $\alpha$ has to be
periodic with unit period. The component $\mathcal{A}_t^+$ is fixed by the Chern Simons equations of motions. 
The fact that the
Y-field needs to have a non-trivial monodromy around the defect follows 
when the BPS condition, which imposes holomorphicity, is combined with the requirement of scale invariance. This configuration breaks R-symmetry to $SU(3) \times U(1)_R$.  The conserved  bosonic symmetries are  now $SU(1,1) \times U(1)_d \times SU(3)$ where $U(1)_d$ is a diagonal combination of space-time and R-symmetry whereas the full supersymmetry group is $SU(1,1|3)$ corresponding to 12 remaining supercharges out of 24.

\paragraph{1/3 BPS:}
 In this case two complex scalars are non-vanishing  and the full solution reads
  
\begin{equation}
\mathcal{Y}^1=\frac{\beta_1}{\sqrt{z}}, \hspace{0.5cm} \mathcal{Y}^2=\frac{\beta_2}{\sqrt{\bar{z}}}, \hspace{0.5cm} \mathcal{A}_z^+= -i \frac{\alpha}{2kz},
\hspace{0.5cm}
\mathcal{A}_t^+= -\frac{1}{|z|}\,(|\beta_1|^2-|\beta_2|^2).
\end{equation}
Again the parameters $\alpha$, $\beta_1$ and $\beta_2$ are free whereas $\mathcal{A}_t^+$ is a derived quantity. Only the relative
phase between $\beta_1$ and $\beta_2$ has physical meaning, and again $\alpha$ can be taken to be real. Hence, the
solution has four real parameters. The remaining R-symmetry is now $SU(2) \times U(1)_R$ and the full remaining bosonic
symmetry $SU(1,1)\times SU(2)\times U(1)_{d'}$ where again $U(1)_{d'}$ is a diagonal combination of 
space-time and R-symmetry. The remaining supergroup invariance is $SU(1,1|2)$ corresponding to eight remaining
supercharges out of 24.\footnote{For $k=1,2$  enhancement of supersymmetry makes this configuration 1/2 BPS. There are also subtleties related to supersymmetry enhancement for the 1/6 BPS configuration. We refer to~\cite{Drukker:2008jm} for details.}

\paragraph{1/6 BPS:} 
All four complex scalars are non-vanishing and the full solution reads
\begin{align}
\mathcal{Y}^1&=\frac{\beta_1}{\sqrt{z}}, \hspace{0.5cm} \mathcal{Y}^2=\frac{\beta_2}{\sqrt{{z}}}, \hspace{0.5cm} 
\mathcal{Y}^3=\frac{\beta_3}{\sqrt{\bar{z}}}, \hspace{0.5cm} \mathcal{Y}^4=\frac{\beta_4}{\sqrt{\bar{z}}}, \hspace{0.5cm} \mathcal{A}_z^+= -i \frac{\alpha}{2kz},\\
\mathcal{A}_t^+&= -\frac{1}{|z|}\,(|\beta_1|^2+|\beta_2|^2-|\beta_3|^2-|\beta_4|^2).
\end{align}
The bosonic symmetry  conserved by this solution is $SU(1,1) \times U(1)_{\hat{d}}$ where $U(1)_{\hat{d}}$ is a diagonal combination of space-time and R-symmetry.
The full supergroup invariance is $SU(1,1|1)$. For $U(1)\times U(1)$ gauge symmetry this solution can actually be rotated into
the solution above and is thus 1/3 BPS. However, the solution admits a non-Abelian generalization, which we will describe shortly and which is 1/6 BPS. For all the symmetry breaking cases $\mathcal{A}^-_\mu = 0$.

 All the vortex configurations described above admit non-Abelian generalizations, in which the $U(N)\times \hat{U}(N)$ gauge symmetry of the parent theory is broken by the vortex background to a Levi-type subgroup.
More precisely, for the 1/2 BPS case one can take
\begin{equation}\label{nonAbelian}
\mathcal{Y}^1=\frac{1}{\sqrt{z}}\begin{pmatrix}
\mathbf{1}_{N_0} \otimes 0 & 0& \cdots & 0\\
0& \mathbf{1}_{N_1}\otimes \beta^{(1)} & \cdots & 0 \\
\vdots & \vdots  & \ddots & \vdots \\
0 & \cdots&  0 & \mathbf{1}_{N_M} \otimes \beta^{(M)}
\end{pmatrix}, \hspace{0.5cm}
\end{equation}
\begin{equation}
\mathcal{A}_z^+
=
-\frac{i}{2kz}
\begin{pmatrix}
\mathbf{1}_{N_0} \otimes 0& 0 & \cdots & 0 \\
0 &  \mathbf{1}_{N_1} \otimes \alpha^{(1)}& \cdots & 0 \\
\vdots & \vdots & \ddots & \vdots \\
0 & 0 & \cdots &  \mathbf{1}_{N_M}\otimes \alpha^{(M)} 
\end{pmatrix}, \hspace{0.5cm}
\mathcal{A}_t^+=- \mathcal{Y}^1 \mathcal{Y}_1^\dagger,
\end{equation}
where $\sum_{i=0}^M N_i=N$. This construction breaks the gauge symmetry to $U(N_0)\times \hat{U}(N_0) \times \, U(N_1)\times \ldots \times U(N_M)$. As in the Abelian case both the $\alpha$'s and
the $\beta$'s can be taken to be real and the $\alpha$'s must be periodic with unit period. The solution is thus described in terms of $2M$ real parameters. In the same way the
1/3 BPS configuration can be generalized to the non-Abelian case by promoting its $\beta_1$, $\beta_2$ and $\alpha$ to diagonal $N\times N$ matrices with blocks of identical 
eigenvalues of multiplicities $N_i$, $i=0,\ldots, M$ where $\sum_{i=0}^M N_i=N$. In this case the relative angles between $\beta_1^{(i)}$ and $\beta_2^{(i)}$ add extra free 
parameters to the description so that this solution is described in terms of $4M$ real parameters. In addition, the temporal component of the gauge field is determined to be
\begin{equation}
\mathcal{A}_t^+=- (\mathcal{Y}^1 \mathcal{Y}_1^\dagger-\mathcal{Y}^2 \mathcal{Y}_2^\dagger).
\end{equation}
The $1/6$-BPS configuration admits an analogous non-Abelian generalization with $8M$ real parameters. For each block $(m)$, the four complex parameters $\beta_I^{(m)}$ are defined modulo a common overall phase, leaving seven real parameters, while the real periodic parameter $\alpha^{(m)}$ provides one additional parameter. Finally the temporal component of the gauge field is fixed as 
\begin{equation}
\mathcal{A}_t^+=- (\mathcal{Y}^1 \mathcal{Y}_1^\dagger+\mathcal{Y}^2 \mathcal{Y}_2^\dagger-
\mathcal{Y}^3 \mathcal{Y}_3^\dagger-\mathcal{Y}^4 \mathcal{Y}_4^\dagger).
\end{equation}

\section{One-point functions of non-protected operators\label{onepoint}}

The requirement of scale invariance of the BPS configuration ensures that the quantum field theory equipped with the resulting monodromy defect
 constitutes a defect CFTs. As is well-known, in a dCFT novel types of conformal
data appear compared to standard CFTs. One such set of data are the one-point function coefficients of conformal operators, $C_{\cal O}$, appearing as
\begin{equation}
\langle {\cal O}_\Delta(x) \rangle = \frac{C_{\cal O}}{r^\Delta},
\end{equation}
where $\Delta$ is the conformal dimension of the operator and  $r$ is the distance to the defect.\footnote{In case there is no rotational invariance
in the transverse plane, ${\cal C}_{\cal O}$ can depend on angles.}
In a semi-classical approximation correlation functions 
of operators in the presence of the defect can be determined simply by inserting the BPS solutions for the fields in the operators. We shall be interested in non-protected gauge invariant local operators in the form of single trace operators, the reason being that these can be efficiently treated by tools of integrability. Such operators are necessarily built
from a string of complex scalar fields alternating between $Y$'s and $Y^\dagger$'s, i.e.
\begin{equation}
\mathcal{O}_{}^{A}
=
C^{(A)\,J_1\cdots J_{L}}_{\mbox{ } I_1\cdots I_{L}}\,
\operatorname{Tr}
\left(
Y^{I_1} Y_{J_1}^{\dagger}
\cdots
Y^{I_{L}} Y_{J_{L}}^{\dagger}
\right).
\label{eq:chiral-primary}
\end{equation} 
 One-point functions of such operators are single valued when taken around the defect. The same must be the case for all other gauge invariant observables in order for the defect to be acceptable as a physical object. 

A class of operators which
plays a particular role in supersymmetric theories is chiral primary operators whose conformal dimensions do not get quantum corrections. Single trace chiral primaries in ABJM theory are obtained by requiring the tensor $C^{(A)\,J_1\cdots J_{L}}_{\mbox{ } I_1\cdots I_{L}}$ above to be symmetric in lower as well as upper indices and traceless under contraction of one upper with one lower index. It was pointed out in~\cite{Drukker:2008jm}, that
in addition to the neutral single-trace chiral primaries, ABJM theory contains baryonically charged chiral primary operators, involving unequal numbers of $Y$ and $Y^\dagger$ fields which must be rendered gauge invariant by an appropriate monopole dressing. We shall not consider such operators here.
Single trace chiral primary operators in ABJM theory are in one-to-one correspondence with spherical harmonics on $S^7/\mathbb{Z}_k$. In order for such an operator
to have a non-trivial one-point function in the background defined by one of the supersymmetric monodromy defects it has to be invariant under the R-symmetry conserved by the defect. This criterion picks out only one spherical harmonic for each value of $L$. The field theory realization of the operator can be written down on a case by case basis and the semi-classical one-point function determined by inserting the appropriate classical fields. This strategy was followed in~\cite{Drukker:2008jm} which computed the one-point function of  a number of chiral primaries for the 1/2 BPS defect.
 
 We shall be concerned with the semi-classical computation of one-point functions of non-protected operators where tools of integrability allow us to obtain the result in a single closed formula  valid for any single trace conformal operator built from scalars. The key point is that in the planar limit of the field theory, integrability allows us to identify any given conformal non-protected single trace operator with a Bethe eigenstate of an integrable spin chain~\cite{Minahan:2009te}.
 More precisely, a conformal single-trace operator of the above form, containing $L$ fields $Y$ and $L$ fields $Y^\dagger$, can be identified at the leading perturbative order (two loops) with an eigenstate of an integrable alternating $SU(4)$ spin chain of length $2L$ and the operator's anomalous conformal dimension at this order then equals the corresponding eigenvalue. The even and odd sites of the chain transform in the fundamental and anti-fundamental representations of $SU(4)$, respectively, with the four possible states at each site corresponding to the four flavours of the $Y$ and $Y^\dagger$ fields. The 
spin chain Hamiltonian takes the form~\cite{Minahan:2008hf}
 \begin{equation}
H
=
\frac{\lambda^2}{2}
\sum_{l=1}^{2L}
\left(
2
-2P_{l,l+2}
+P_{l,l+2}K_{l,l+1}
+K_{l,l+1}P_{l,l+2}
\right),
\label{eq:scalar-mixing-matrix}
 \end{equation}
 with $P$ the permutation operator and $K$ the trace, 
 and it acts on a spin chain state as the quantum field theory's two-loop dilatation operator would act on the corresponding single trace operator.  Here, we are considering the 
 planar, 't Hooft limit  $N\rightarrow \infty, k\rightarrow \infty$ with $\lambda=\frac{N}{k}$ kept fixed and the expression above is the leading order contribution in $\lambda$.\footnote{In particular, considerations pertaining to $k=1,2$ are irrelevant for the following discussion.}

\begin{figure}
 \begin{center}
 \includegraphics[width=3.2cm] {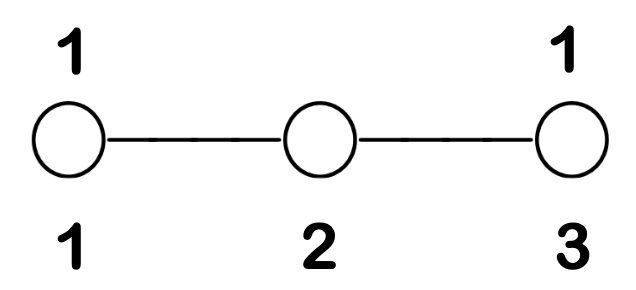}
 \end{center}
\caption{\label{SU4Dynkin}The Dynkin diagram of $SU(4)$ . Above the nodes are the weights of the relevant spin representation while the numbers below the nodes are the labels of the $Q$-functions.}
\end{figure}

The Bethe eigenstates of the Hamiltonian are highest weight states of $SU(4)$ multiplets and thus in particular have three Dynkin labels associated with them. In the framework of the algebraic Bethe ansatz the eigenstates are described as excitations on top of a chosen spin chain vacuum and are characterized in terms of three sets of Bethe rapidities, one for each of the 
three nodes of the Dynkin diagram of $SU(4)$,
\begin{equation}
\{u_{i}^{(1)}\}_{i=1}^{r_1}, \hspace{0.5cm}\{u_{j}^{(2)}\}_{j=1}^{r_2},  \hspace{0.5cm}\{u_{k}^{(3)}\}_{k=1}^{r_3},
\end{equation}
with the node numbering appearing below the Dynkin diagram in figure~\ref{SU4Dynkin}.

As the vacuum of our spin chain we shall take the state
\begin{equation}
\mbox{Tr}(Y^1 Y_4^{\dagger})^L,
\end{equation}
with the  Dynkin labels $(L,0,L)$ corresponding to the states $Y^1$ and $Y_4^\dagger$ carrying labels $(1,0,0)$ and $(0,0,1)$. We assume that $L>1$.
A Bethe eigenstate with $r_i$ rapidities of type $i$ for $i=1,2,3$, will have  Dynkin labels given by
\begin{equation}
(a_1, a_2, a_3)= (L-2r_1+r_2, r_1-2r_2+r_3, L+r_2-2r_3).
\end{equation} 
Denoting the number of $Y^A$-fields in an operator by $n_A$ and the number of $Y_A^\dagger$-fields by $m_A$, the field
content of the single trace components constituting  a Bethe eigenstate has to fulfil the following constraint
\begin{equation} \label{fieldcounting}
n_1-m_1= L-r_1,\hspace{0.5cm} n_2-m_2=r_1-r_2, \hspace{0.5cm} n_3-m_3=r_2-r_3,\hspace{0.5cm} n_4-m_4=r_3-L,
\end{equation}
where only excitation numbers which give rise to non-negative Dynkin labels are meaningful.

The Bethe roots must obey three sets of coupled Bethe equations with respectively $r_1$, $r_2$ and $r_3$ equations in each set. These equations are expressed in group theoretical language as
\begin{eqnarray}\label{BEs}
&&\left(\frac{u_{j}^{(a)}-\frac{iq_a}{2}}{u^{(a)}_{j}+\frac{iq_a}{2}}\right)^L
 \prod_{bk}^{}\frac{u^{(a)}_j-u^{(b)}_k+\frac{iM_{ab}}{2}}{u^{(a)}_{j}-u^{(b)}_{k}-\frac{iM_{ab}}{2}}
 =\left(-1\right)^{\frac{M_{aa}}{2}},
\end{eqnarray}
where $M_{ab}$ refers to the $SU(4)$ Cartan matrix and $q_a$ to the labels of the elementary spin representation
\begin{equation}
 M=\begin{bmatrix}
 2  & -1 &  0 \\ 
 -1  & 2 & -1  \\ 
  0  & -1 &  2 \\ 
 \end{bmatrix},\qquad 
 q=\begin{bmatrix}
  1 \\ 
  0 \\ 
  1 \\ 
 \end{bmatrix}.
\end{equation}
The  non-vanishing $q_a$-labels appear on top of the nodes of the 
Dynkin diagram in figure~\ref{SU4Dynkin}.
The roots must furthermore fulfil the following momentum constraints in order for the spin chain state to reflect the cyclicity of the
corresponding operator
\begin{equation}
\prod_{j=1}^{r_1} \left( \frac{u^{(1)}_{j}-\frac{i}{2}}{u^{(1)}_{j}+\frac{i}{2}}\right)
\prod_{k=1}^{r_3} \left( \frac{u^{(3)}_{k}-\frac{i}{2}}{u^{(3)}_{k}+\frac{i}{2}}\right)=1.
\end{equation}

The one-point function of a conformal operator at the lowest loop level is evaluated simply by replacing all fields in the
operator with their classical values. This procedure can conveniently
be expressed as computing the overlap between a specially designed matrix product state and the Bethe eigenstate, 
$| {\bf u}\rangle$, representing the operator~\cite{deLeeuw:2015hxa,Buhl-Mortensen:2015gfd}, which in the present
case implies
\begin{equation}\label{O(x)}
 \left\langle \mathcal{O}_L(x)\right\rangle=
 \frac{1}{|z|^L}\,\,
 \frac{1}{\lambda ^LL^{\frac{1}{2}}}\,\,\frac{\left\langle {\rm MPS}\right.\!\left|  {\bf u}\right\rangle}{\left\langle {\bf u} \right.\!\left| {\bf u}  \right\rangle^{\frac{1}{2}}},
\end{equation}
with
\begin{equation}\label{gen-MPS}
 {\rm MPS} _{\hphantom{A}a_2\,\ldots\, a_{2L}}^{a_1\,\ldots\, a_{2L-1}}=\mathop{\mathrm{Tr}}\widetilde{\mathcal{Y}}^{a_1}
\widetilde{ \mathcal{Y}}^\dagger _{a_2}\ldots \widetilde{\mathcal{Y}}^{a_{2L-1}}\widetilde{\mathcal{Y}}^\dagger _{ a_{2L}}.
\end{equation}
where the $\widetilde{\mathcal{Y}}$'s are the classical values of the $Y$'s with all
$z$-dependence scaled away, i.e. $\widetilde{\mathcal{Y}}$ are constant matrices.\footnote{The fact that the $z$-dependence of the one-point function organizes into a pure $1/|z|^L$ factor will be evident from the selection rules that we derive later and which appear in eqns.~(\ref{selection1}),
(\ref{selection2}), and (\ref{selection3}) }

Under certain circumstances a matrix product state can be integrable meaning that its overlaps with Bethe eigenstates
can be found in a closed form. If the symmetry algebra of the spin chain is given by some Lie group $\mathfrak{g}$ and the the matrix product state breaks the symmetry down to a subalgebra $\mathfrak{h}$, a necessary condition for integrability of the
matrix product state is that  $(\mathfrak{g},\mathfrak{h})$ 
constitutes a symmetric pair~\cite{Gombor:2020kgu}.  In our case this criterion is clearly fulfilled. The symmetry algebra of
our spin chain is $\mathfrak{su}(4)$ and the preserved sub-algebra is $\mathfrak{su}(3) \times \mathfrak{u}(1)$ as all three defect set-ups pick out a single vector in $\mathbb{C}^4$.

To prove the integrability of the matrix product state and to extract the desired overlaps one needs to solve
the so-called KT-relation for the reflection matrix, $K$~\cite{DeLeeuw:2019ohp,Gombor:2021hmj,Gombor:2024iix,Gombor:2025wvu}.
The KT relation can be crossed or uncrossed, the type depending on
the nature of the symmetric pair describing the symmetries of the boundary problem. All possible symmetric pairs for the
classical Lie algebras together with the crossing properties of the KT equation were listed in~\cite{Gombor:2024iix}. We
learn from there that our case requires an uncrossed KT-relation which for a general matrix product state takes the form\footnote{In general the argument of the monodromy matrix on the right hand side of eqn.~(\ref{KTrelation}) takes the form
$(-u+\xi)$ where $\xi$ is convention dependent.}
\begin{equation}\label{KTrelation}
    K_{ik}^{\alpha \gamma} (u) \big \langle \mbox{MPS}^{\gamma \beta} \big| T_{kl}(u) = \big \langle \mbox{MPS}^{\alpha \gamma} \big| {T}_{ik}(-u) K_{kl}^{\gamma \beta}(u).
\end{equation}
The quantity $T_{ik}$ is the monodromy matrix and the indices shown live in an auxiliary space.  It is sufficient to solve the KT relation for a spin chain whose length equals the number of fields in an elementary building block of the matrix product state. In the present case, this corresponds to a block of length two, consisting of one site in the fundamental representation and one site in the anti-fundamental representation. The matrix product state occurs in the form where its trace has been cut open and the greek letters are the corresponding free matrix indices. As all the classical fields that we consider are given in terms of diagonal matrices, we can restrict ourselves to studying only matrix product states of bond dimension  one (and obtain our result as a sum over overlaps).\footnote{We note that matrix product states involving diagonal matrices also appeared in the computation of three-point functions of two maximal giant gravitons 
and a single tiny~graviton~\cite{Yang:2021hrl} as well as for the computation of one-point functions in the background of particular classes of supersymmetric Wilson loops~\cite{Jiang:2023cdm}. We will explain the precise relation later.}
Once we restrict ourselves to such states, we
have $c$-numbers instead of matrices and the greek indices become obsolete, i.e. we can replace $(\widetilde{\mathcal{Y}}^{a})^{\rho \sigma} = \beta_a \mathds{1}^{\rho \sigma}_{1 \times 1} = \beta_a$. With this, our (bond dimension one) matrix product state becomes
\begin{equation}
\langle \text{MPS}
|= \sum_{a,b=1}^4 \bra{a}  \bra{\bar{b}}\, {\beta}_a  {\beta}_b^*. 
\end{equation}
In explicit terms the monodromy matrix is thus replaced by
\begin{equation}
T_{kl;\:acbd}(u) = \hat{\mathcal{L}}_{kj;\: {b}{d}}(u) \mathcal{L}_{jl; \:ac}(u),
\end{equation}
where the Lax operators ${\mathcal{L}}$ and $\hat{\mathcal{L}}$ are explicitly known~\cite{Gombor:2024iix}
\begin{subequations}
    \begin{align}
        \mathcal{L}_{kl; \: bd}(u) &= \sum_{i,j=1}^4 (e_{ij})_{kl} \otimes \left( (\delta_{ij})_{bd}+\frac{1}{u-1} (e_{ji})_{bd}\right),  \\
        \hat{\mathcal{L}}_{lj; \: ac}(u) &= \sum_{p,q=1}^4 (e_{pq})_{lj} \otimes \left( (\delta_{pq})_{ac}+\frac{1}{-u-1} (e_{pq})_{ac}\right). 
    \end{align}
\end{subequations}
and our task is to find a $4\times 4$ reflection matrix $K_{ij}$ which acts in the auxiliary space which 
we have taken to be the fundamental one.
Writing out all terms one finds that the following reflection matrix solves the KT-relation
\begin{equation}
K_{ij} = \delta_{ij}+ \frac{2u}{1-u} \frac{\beta_i^* \beta_j}{|\beta|^2}. \label{K-matrix}
\end{equation}
Where we defined $|\beta|^2 = |\beta_1|^2 + |\beta_2|^2+|\beta_3|^2+|\beta_4|^2$. We stress that this solution includes all cases, 1/2--BPS, 1/3--BPS and 1/6--BPS defects, with only difference being the number of $\beta$'s that are non-zero. This K-matrix also very
recently appeared in~\cite{Bai:2026hun} which performed a systematic search
for integrable matrix product states built from blocks of site-length one to four.

Knowing the reflection matrix it is  possible to extract the overlap formula for the corresponding MPS state by
a recursive procedure~\cite{Gombor:2024iix,Gombor:2025wvu}.  But already the symmetries of the problem give rise to
strict selection rules for the overlaps. 
First of all, overlaps are only non-trivial if the Bethe roots come 
in pairs with opposite signs which is a consequence of the matrix product state being annihilated by all the 
odd charges of the spin chain, a property which follows when one multiplies both sides of the KT relation
by the inverse of the $K$-matrix and takes the trace. We notice from eqn.~(\ref{K-matrix}) that the K-matrix is indeed
invertible. 
The pairing of Bethe roots can be realized in two distinct ways, referred to as chiral and achiral pairing. In the chiral case, roots are paired within each set of Bethe roots associated with a given node of the Dynkin diagram. By contrast, in the achiral case, roots associated with different nodes are paired with one another. The chirality of the overlap is determined by the corresponding symmetric pair $(\mathfrak g,\mathfrak h)$. According to the classification of \cite{Gombor:2024iix}, the symmetric pair $(\mathfrak{gl}_4,\mathfrak{gl}_3\oplus\mathfrak{gl}_1)$ relevant here gives rise to an achiral pairing: Bethe roots associated with the two outer nodes of the Dynkin diagram are paired with each other, while the roots associated with the middle node are paired among themselves, i.e
\begin{equation}
     \{u^{(1)}_j\}_{j=1}^{r_1} = \{ -u^{(3)}_k\}_{k=1}^{r_3} , \hspace{0.7cm} \{u^{(2)}_j\}_{j=1}^{r_2} = \{ -u^{(2)}_j\}_{j=1}^{r_2}.
\end{equation}
In particular, this implies that the number of roots at the two outer nodes of the Dynkin diagram have to be the same
\begin{equation}
r_1=r_3. \label{selection1}
\end{equation}
The general overlap formula looks as follows~\cite{Gombor:2024iix}
\begin{equation}
    \frac{\bra{\mbox{MPS}} \mathbf{u}\rangle}{\sqrt{\bra{\mathbf{u}} \mathbf{u}\rangle}} = \left[\sum_{l=1}^{d_B} A_l \prod_{\nu=1}^{n_+}\prod_{k=1}^{r_\nu^+}  {\widetilde{\mathcal{F}}}^{(\nu)}_l(u_{k}^{(\nu)})\right] \times \sqrt{S\det G},
\end{equation}
where $S\det G$ is the superdeterminant of the Gaudin matrix~\cite{Kristjansen:2020vbe},  a quantity which is independent of the matrix product state, and where $r_\nu^+$ is the number of positive roots associated with a given node. For a Dynkin node where
the roots are paired with the roots of another node $r_\nu^+=r_\nu$. If the roots are paired within the same node the product is over positive roots and $r_\nu^+= r_\nu/2 $. Which roots are denoted as positive is not important since the associated
${\widetilde{\mathcal F}}$-functions are invariant under
$u\leftrightarrow -u$. In practice, the explicit dependence on the Bethe roots in the product over $\widetilde{\mathcal{F}}$'s always
organizes into products of Baxter polynomials, see e.g.\ the review \cite{Kristjansen:2024dnm}.
The quantity $d_B$ is the bond dimension of the matrix product state 
and $n_+$ is the number of Dynkin nodes which carry independent sets of Bethe roots, in our case
\begin{equation}
d_B=1, \hspace{0.5cm} n_+=2.
\end{equation}
Hence, we shall leave out the index $l$ on our quantities. 
The pre-factor $A$ is given by
\begin{equation}
A = \langle \mbox{MPS} | \Omega\rangle,
\end{equation}
where $|\Omega\rangle$ is the vacuum state of the spin chain. The $\widetilde{\mathcal{F}}$'s
can be determined from the reflection matrix $K_{ij}$. To do this, one needs to define a set of nested $K$-matrices, which for us terminates at the second level of nesting. One derives a $K$-matrix at the first nested level as
\begin{equation}
K_{ij}^{(2)}(u)=K_{ij}(u)-K_{i1}(u) \left[K_{41}(u)
\right]^{-1} K_{4j}(u),  \hspace{0.5cm} i,j=2,3.
\end{equation}
In the next step one introduces $G$-functions as 
\begin{equation}
G^{(1)}(u)= K_{41}(u), \hspace{0.5cm} G^{(2)}(u) = K^{(2)}_{32}(u), \hspace{0.5cm}
G^{(3)}(u)= K_{23}^{(2)}-K_{22}^{(2)} \left[ K_{32}^{(2)}(u)\right]^{-1} K_{33}(u).
\end{equation}
For a generic matrix product state all of these objects would have indices in the boundary space and hence be $d_B\times d_B$ matrices.
From that perspective our case is considerably simpler as $d_B=1$ and we are simply dealing with $c$-numbers.
 In a next step one computes  ${\cal F}$-functions which are then also simply ratios of $c$-numbers, more precisely
\begin{equation}
 \mathcal{F}^{(1)} (u) = \frac{G^{(2)}(u)}{G^{(1)}(u)}, \hspace{0.5cm}
     \mathcal{F}^{(2)} (u) = \frac{G^{(3)}(u)}{G^{(2)}(u)}.
\end{equation}
From here one finally reads~\cite{Gombor:2025wvu}
\begin{align}
 \widetilde{\mathcal{F}}^{(1)} (u) &=  \mathcal{F}^{(1)}\left( iu+\frac{1}{2}\right),\\
 \widetilde{ \mathcal{F}}^{(2)} (u) &=  \mathcal{F}^{(2)}(iu) \sqrt{\frac{u^2}{u^2+\frac{1}{4}}}.
\end{align}
However, we encounter some subtleties in the implementation of this procedure. A minor problem is that $K_{41}(u)=0$ for
the $1/2$-BPS and the $1/3$-BPS set-up where $\beta_4=0$. Such a problem can generally be dealt with by performing 
$\mathfrak{gl}(4)$ rotation of the vacuum state, but we shall simply solve the general $1/6$-BPS case and obtain the others
by sending certain of the parameters to zero. A more complicated issue is that the element $K_{32}^{(2)}$ appears to be zero for all choices of the $\beta$-parameters. Namely, one finds for the nested $K$-matrix
\begin{equation}
    K_{ij}^{(2)}(u) = 
    \begin{pmatrix}
        1 & 0 \\
        0 & 1
    \end{pmatrix},
\end{equation}
where the indices can take the values $i,j=2,3$. There exists a resolution to this type of problem which is known as a 
performing a deformation of the K-matrix~\cite{Gombor:2025wvu}. This introduces a number of regularization parameters which must be sent to zero at the end of the computation. If one can decompose the  $K$-matrix found as a solution to the KT relation in the form 
\begin{equation} \label{twistedKmatDecomp}
   K_{ij}(u) = \sum_{k=1}^{N} L'_{ik}(u) L_{kj}(-u),
\end{equation}
where $L$ and ${L}'$ satisfy:
\begin{equation}\label{L'def}
    \sum_{k=1}^{N} {L}_{ik}'(u) {L}_{kj}(u) = \delta_{ij},
\end{equation}
then the deformed $K$-matrix, $\widetilde{K}$, given by
\begin{equation}
    \widetilde{{K}}_{i,j}(u|\mathfrak{b}) ={K}_{ij}(u) + u \sum_{k=1}^{N/2} \mathfrak{b}_k  {L}'_{i, N+1-k}(u) {L}_{kj}(-u),
\end{equation}
will fulfil the same reflection equation as the undeformed one and the singularities preventing the determination of the overlap will have been regularized by means of the free parameters $\mathfrak{b}_i$. We can implement this procedure on the matrix
$K_{ij}^{(2)}$ above with both $L$ and $L'$ equal to the identity matrix.
This gives the following regularized $K$-matrix
\begin{equation}
    \widetilde{K}^{(2)}(u) = 
    \begin{pmatrix}
        1 &  \, 0 \, \\
        \mathfrak{b}_1 u & \, 1\,
    \end{pmatrix}.
\end{equation}
With this, the operators $G$ become:
\begin{equation}
    G^{(2)}(u) =\widetilde{K}^{(2)}_{32}(u)= \mathfrak{b}_1 u,
\end{equation}

\begin{equation}
    G^{(3)}(u) = \widetilde{K}^{(2)}_{23}(u) - \widetilde{K}^{(2)}_{22}(u) \frac{1}{\widetilde{K}^{(2)}_{32}(u)} \widetilde{K}^{(2)}_{33}(u) = -\frac{1}{\mathfrak{b}_1 u}.
\end{equation}
This allows us to compute the $\mathcal{F}$-functions 
\begin{equation}
    \mathcal{F}^{(1)}(u) = \frac{G^{(2)}(u)}{G^{(1)}(u)} = \mathfrak{b}_1 \frac{(1-u)|\beta|^2}{2 (\beta_1  \beta_4^*)},
\end{equation}
\begin{equation}
    \mathcal{F}^{(2)}(u) = \frac{G^{(3)}(u)}{G^{(2)}(u)} = -\frac{1}{\mathfrak{b}_1^2 \; u^2},
\end{equation}
and therefore the $\widetilde{\mathcal{F}}$  become
\begin{equation}
    \widetilde{\mathcal{F}}^{(1)}(u) = \mathfrak{b}_1 \frac{-i(u - i/2)}{2} \frac{|\beta|^2}{(\beta_1  \beta_4^*)}, \label{Ftilde1}
\end{equation}
\begin{equation}
    \widetilde{\mathcal{F}}^{(2)}(u) = \frac{1}{\mathfrak{b}_1^2} \sqrt{\frac{1}{u^2(u^2+1/4)}}.
\end{equation}
Finally, we find
\begin{equation}
\langle \mbox{MPS} | \Omega \rangle= {(\beta_1\beta_4^*)}^L.
\end{equation}
Collecting everything we get
\begin{eqnarray}
    \frac{\bra{\mbox{MPS}} \mathbf{u}\rangle}{\sqrt{\bra{\mathbf{u}} \mathbf{u}\rangle}} &=& 
     (\beta_1 \beta_4^*)^{L-r_1} \mathfrak{b}_1^{r_1-r_2} |\beta |^{2r_1} 2^{-r_1} (-i)^{r_1} \times \\    
   & &  \prod_{j=1}^{ r_2/2} \frac{1}{\sqrt{\left(u_j^{(2)}\right)^2
    \left(\left(u_j^{(2)}\right)^2+\frac{1}{4}\right)}}
    \prod_{k=1}^{r_1}\left(u_k^{(1)}+\frac{i}{2}\right)
    \times \sqrt{S\det G}.
\end{eqnarray}
We thus see that the requirement that the result be independent of the regularization parameter enforces the additional
selection rule (on top of the rule $r_1=r_3$ deriving from the fact that the overlap should be achiral)
\begin{equation}
r_1=r_2. \label{selection2}
\end{equation}
After implementing this rule we get the following result for the overlap
\begin{equation} \label{overlap}
    \frac{\bra{\mbox{MPS}} \mathbf{u}\rangle}{\sqrt{\bra{\mathbf{u}} \mathbf{u}\rangle}} =
    (\beta_1 \beta_4^*)^{L-r_1}  |\beta |^{2r_1} 2^{-r_1} (-i)^{r_1}       \frac{ Q_1\left(\frac{i}{2}\right)}{\sqrt{Q_2(0) Q_2\left(\frac{i}{2}\right)}}
    \times \sqrt{S\det G},
\end{equation}
where $Q_i(u)$ is the Baxter polynomial associated with the node $i$, i.e.\ $Q_i(u)=\prod_{j=1}^{r_i}(u-u_j^{(i)})$.
From here we learn that for the $1/2$-BPS and for the $1/3$-BPS case where $\beta_4=0$ we get an additional
selection rule
\begin{equation}
L=r_1 \hspace{0.5cm} \mbox{for} \hspace{0.5cm} \beta_4=0. \label{selection3}
\end{equation}
To get the one-point function of the corresponding operator we just need to supplement the overlap formula with the
prefactor given in eqn.~(\ref{O(x)}). We notice as expected that  the space-time dependence always amounts to a  factor of
$\frac{1}{|z|^L}$ which follows from our selection rules and the 
relation~(\ref{fieldcounting}).
The overlap formula~(\ref{overlap}) has been derived following the procedure 
of~\cite{Gombor:2024iix} where it is assumed that there are no unpaired zero roots and no singular roots, i.e.\ no roots equal to $\pm i/2$. However, it is known how to regularize overlap formulas to account for these special configurations.  The case of an unpaired zero-root for the XXZ spin chain was treated already in~\cite{Brockmann_2014} and a generalization to nested systems was presented in~\cite{deLeeuw:2016umh,DeLeeuw:2018cal,Kristjansen:2021xno}. In this case, one simply leaves out the zero root of the Baxter polynomial in the pre-factor and handles the root in a particular way when computing the super determinant of the Gaudin matrix~\cite{deLeeuw:2016umh,DeLeeuw:2018cal,Kristjansen:2021xno}.
For paired zero roots which can occur for 
achiral configurations no particular care is needed.
 Singular roots require a regularization procedure already at the level of the Bethe equations and one must evaluate the Baxter polynomials and the Gaudin determinant in the overlap formula using the same regularization procedure, and only remove the regulator at the end of the computation~\cite{Kristjansen:2021xno}. 
We thus
expect our overlap formula to be valid for singular Bethe root configurations provided the formula is applied to regularized roots. 
 In Appendix~\ref{check} we confirm the validity of the overlap formula by explicitly computing the one-point function for a number
of short operators. We include both a case with an unpaired zero root and a case
with singular roots.

We remind the reader that eqn.~(\ref{overlap}) was derived for the case of bond dimension one. The result immediately
generalizes to any case where the classical fields are 
diagonal, cf.~eqn.~(\ref{nonAbelian}). The result in that case reads
\begin{align}
&\langle {\cal O}_L(x)\rangle
= \frac{1}{|z|^L}\frac{1}{\lambda^L L^{1/2}}
   \frac{\langle \mathrm{MPS} \vert \mathbf{u}\rangle}
        {\sqrt{\langle \mathbf{u}\vert\mathbf{u}\rangle}}
   \notag\\
&\qquad=\frac{1}{|z|^L}\frac{1}{\lambda^L L^{1/2}} \sum_{k=1}^M N_k 
\left(\beta_1^{(k)}(\beta_4^{(k)})^*\right)^{L-r_1}
|\beta^{(k)}|^{2r_1}2^{-r_1}(-i)^{r_1}
\frac{Q_1\left(\frac{i}{2}\right)}
{\sqrt{Q_2(0)Q_2\left(\frac{i}{2}\right)}}
\sqrt{S\det G}.
\label{overlapfinal}
\end{align}
Let us note that our result includes as a special case the overlap which appears in the computation of the three-point function involving two maximal giant gravitons and
a tiny graviton~\cite{Yang:2021hrl}. In the latter case one needs the overlap between the 
Bethe eigenstates and a matrix product state of bond dimension two, corresponding to a situation where 
$ Y^1$, ${ Y}^4$, ${ Y}_1^\dagger$, ${{ Y}}_4^\dagger$ have non-vanishing $2 \times 2$ matrices associated with them and the matrix product state is again defined as in~(\ref{gen-MPS}). 
The matrices involved are off-diagonal but the products of pairs of matrices which are the only quantities which enter the KT relation (and the only quantities which are relevant for the overlap) are diagonal and of the following form
\begin{eqnarray}
\widetilde{{\mathcal Y}}^1 \widetilde{{\mathcal Y}}_1^\dagger =
\begin{pmatrix}
i &0 \\ 0 &i
\end{pmatrix},&\hspace{0.5cm}&
\widetilde{{\mathcal Y}}^4 \widetilde{{\mathcal Y}}_4^\dagger =
\begin{pmatrix}
i &0 \\ 0 &i
\end{pmatrix}, \hspace{0.5cm}
\widetilde{{\mathcal Y}}^1 \widetilde{{\mathcal Y}}_4^\dagger =
\begin{pmatrix}
-i &0 \\ 0 &i 
\end{pmatrix},\hspace{0.5cm}
\widetilde{{\mathcal Y}}^4 \widetilde{{\mathcal Y}}_1^\dagger = 
\begin{pmatrix}
-i &0\\ 0 &i
\end{pmatrix}
\end{eqnarray}
This case is of the type covered by the formula~(\ref{overlap})  with
\begin{equation}
M=2, \hspace{0.5cm}N_1=N_2=1, \hspace{0.5cm}\beta_1^{(1)} (\beta_4^{(1)})^*= -i,
\hspace{0.5cm}
\beta_1^{(2)} (\beta_4^{(2)})^*= i,
\end{equation}
Furthermore, we can make the  replacement
\begin{equation}
|\beta^{(1)}|^2=\beta_1^{(1)}(\beta_1^{(1)})^*+\beta_2^{(1)}(\beta_2^{(1)})^*
+\beta_3^{(1)}(\beta_3^{(1)})^*+\beta_4^{(1)}(\beta_4^{(1)})^*
\rightarrow 2 i, \hspace{0.5cm} |\beta^{(2)}|^2\rightarrow 2i,
\end{equation}
as we have nowhere in our derivation assumed $|\beta^{(k)}|^2$ to be real. This
results in the following overlap formula
\begin{equation}\label{graviton}
 \frac{\langle \mathrm{MPS} \vert \mathbf{u}\rangle}
        {\sqrt{\langle \mathbf{u}\vert\mathbf{u}\rangle}}=
        \left(  (-i)^{L-r_1}+i^{L-r_1}\right)
        \frac{Q_1\left(\frac{i}{2}\right)}
{\sqrt{Q_2(0)Q_2\left(\frac{i}{2}\right)}}
\sqrt{S\det G}.
\end{equation}
In~\cite{Yang:2021hrl} a very similar formula was presented, the only difference being that instead of $Q_1(i/2)$ the quantity $\sqrt{\prod_{j=1}^{r_1} ((u_j^{(1)})^2+\frac{1}{4})}$ appeared. The two,
however, agree up to   a phase if one takes into account that the roots, if not real, come in complex conjugate pairs.\footnote{In \cite{Yang:2021hrl}, the
exponent of the two factors of $(i)$ and $(-i)$ was written as 
$L-\mbox{(no of 4 on odd sites)}-\mbox{(number of $\bar{1}$ on even sites)}$ which 
is exactly $L-r_1$. The factor $2^{\Delta-J}$ in their formula is a kinematical
factor which is not part of the overlap.} 
In reference~\cite{Yang:2021hrl} the
formula was found by combining arguments based on the coordinate space Bethe ansatz with numerical
investigations and a relatively complicated phase factor expressed in terms of
Bethe roots had to be chosen to get as simple as possible
a result. Our approach is based on the algebraic Bethe ansatz via the KT relation and immediately gives a simple result expressible in terms of $Q$-functions. The phase in our case is determined by the phase of the creation
operators, typically denoted as $B$'s, which are used to form the eigenstate
in the algebraic approach. An overlap formula very similar to 
eqn.~(\ref{graviton}) also appeared in the computation of scalar one-point functions in the background of a selected set of supersymmetric Wilson loops in ABJM theory~\cite{Jiang:2023cdm}. It was derived by a recursive 
strategy which was a precursor of (and equivalent to) the one used here.
The situations considered there correspond in our language to
\begin{equation}
\beta_k\ \beta_j^* \rightarrow 
\text{diag}(i,i,i,-i)\hspace{0.3cm} \mbox{or} \hspace{0.3cm} 
\text{diag}(i,-i,-i,-i)\hspace{0.3cm}\mbox{or}\hspace{0.3cm}\text{diag}(i,i,-i,-i).
\end{equation}
The two first of these are easily recovered by our formula~(\ref{overlap}), where we first of
all notice that we need $r_1=L$ and secondly that $|\beta|^2\rightarrow \pm 2i$.   This leads to the overlap formula
\begin{equation}\label{Wilson}
 \frac{\langle \mathrm{MPS} \vert \mathbf{u}\rangle}
        {\sqrt{\langle \mathbf{u}\vert\mathbf{u}\rangle}}=(\pm 1)^L
        \frac{Q_1\left(\frac{i}{2}\right)}
{\sqrt{Q_2(0)Q_2\left(\frac{i}{2}\right)}}
\sqrt{S\det G},
\end{equation}
which up to a phase which agrees with the result of~\cite{Jiang:2023cdm}. The third case is 
not immediately covered by our formula, but would require a further regularization, as it implies $|\beta|^2\rightarrow 0$.

\section{Towards quantization \label{quantization}}

The BPS solutions constitute special classical configurations of the fields and as such give rise to the classical expressions for one-point functions which we proved to be integrable above. A natural question is what happens at the quantum level. In this
section we initiate the discussion of this question. In order to quantize around the defect background one expands all
fields around their classical values, i.e.\
\begin{equation}
    Y^A = \mathcal{Y}^A + \widetilde{Y}^A , \hspace{0.5cm}
    A = \mathcal{A} + \widetilde{A}, \hspace{0.5cm}
     \hat{A} = \hat{\mathcal{A}} + \widetilde{\hat{A}}.
   \end{equation}
In the following we will drop the tildes on the quantum fields for ease of notation. As usual in such an expansion, terms linear in the fluctuation fields vanish by the equations of motion and the BPS conditions.  At the quadratic level, the background introduces a mixing between scalar and gauge fields which has to be disentangled in order to set up the perturbative program. While fermions in general acquire mass terms as a result of the background they do not mix with other fields and we will mostly leave
the fermions out of our considerations. Even for the bosons alone the mixing problem is in general extensive. For instance the interaction
potential for the $Y$-fields, given by the two last lines in equation~(\ref{ABJM action}), gives rise to three classes of quadratic
(mass-like) mixing contributions
\begin{equation}
m^2_{Y^A Y^\dagger_A} + m^2_{Y^\alpha Y^\dagger_\beta} + m^2_{YY},
\end{equation}
where the contributions are split according to their index mixing structure. In particular, the last term contains mixing terms of the type
$YY$ and $Y^\dagger Y^\dagger$.  The three mass mixing contributions were worked out explicitly in~\cite{Kristjansen:2024zvl} and consist of respectively 6, 10 and 12 terms. In order not to clutter the discussion with extensive book-keeping we shall
restrict ourselves to considering only BPS configurations where all classical fields are proportional to the unit matrix, i.e.
\begin{equation}
\mathcal{Y}^A= \beta^A \, \cdot \mathds{1}_{N\times N}. \label{background}
\end{equation}
In this case
all terms above cancel out.
In appendix~\ref{general case} we briefly discuss the expanded action for the general case and sketch the strategy of its treatment.
Here, we are thus left with the expansion of the kinetic term for the scalars and the Chern-Simons terms for the gauge fields, 
i.e.\
the first line of eqn.~(\ref{ABJM action}). The kinetic part for the scalars gives rise to the following terms
\begin{eqnarray}
 \Tr  D_\mu Y^\dagger_A D^\mu Y^A &\rightarrow & \Tr \left( \partial_\mu Y^\dagger_A \partial^\mu Y^A +  \frac{|\beta|^2}{|z|} (A^\mu A_\mu + \hat{A}^\mu \hat{A}_\mu - 2 A_\mu \hat{A}^\mu)\right. \\
   &&+i \mathcal{Y}^\dagger_A (\hat{A}^\mu - A^\mu) \partial_\mu Y^A + i \partial_\mu \mathcal{Y}_A^\dagger (A^\mu - \hat{A}^\mu)+ \\
    && \left.+ i (\hat{A}_\mu-A_\mu) Y^\dagger_A \partial^\mu \mathcal{Y}^A + i \mathcal{Y}^A(A_\mu - \hat{A}_\mu) \partial^\mu Y^\dagger_A\right),
   \end{eqnarray}
whereas the Chern-Simons term, ${\cal L}_{CS}$, contributes to the quadratic Lagrangian with
\begin{equation}
\mathcal{L}_{CS}\rightarrow
\Tr \varepsilon^{\mu\nu\lambda} (-A_\mu \partial_\nu A_\lambda - 2 i A_\mu A_\nu \mathcal{A}_\lambda + \hat{A}_\mu \partial_\nu \hat{A}_\lambda - 2i \hat{A}_\mu \hat{A}_\nu \hat{\mathcal{A}}_\lambda).
\end{equation}
Notice that the terms of the $A_\mu A_\nu \mathcal{A}_\lambda$ cancel out because they are contracted with the $\varepsilon$-tensor \label{symmetry} and the fields can be freely exchanged because of the trace and the fact that the background field $\mathcal{A}$ is proportional to the identity. Introducing $A^{\pm}= A\pm \hat{A}$ we can write the resulting quadratic bosonic action as 
\begin{equation}
    \begin{split}
    & \mathcal{L}_{bos} = \frac{k}{4\pi} \Tr \biggl( - \frac{\varepsilon^{\mu\nu\lambda}}{2} (A^+_\mu \partial_\nu A_\lambda^- + A_\mu^- \partial_\nu A_\lambda^+)  + \\
    & + \frac{|\beta|^2}{|z|^2} A_\mu^- A^{- \mu} + \partial_\mu Y_A^\dagger \partial^\mu Y^A -  \\
    & -i \mathcal{Y}_A^\dagger A^{-\mu} \partial_\mu Y^A + i \partial_\mu \mathcal{Y}_A^\dagger \cdot A^{-\mu} Y^A - i \partial_\mu \mathcal{Y}^A \cdot A^{- \mu} Y^\dagger_A + i \mathcal{Y}^A A^{- \mu} \partial_\mu Y^\dagger_A
    \biggr),
    \end{split}
\end{equation}
where we notice that only $A^-$ acquires a mass term and only $A^-$ couples directly to the scalars. We gauge fix by adding
the following term plus its associated ghost terms to the action
\begin{equation}
    \mathcal{L}_{gf} = \frac{k}{4 \pi} \Tr \biggl( - \frac{1}{2 \xi} G^2 + \frac{1}{2 \xi} \hat{G}^2 \biggr),
\end{equation}
where $G$ and $\hat{G}$ are given by
\begin{equation}
    G = \eth_\mu A^\mu + i \xi \biggl( Y^A \mathcal{Y}_A^\dagger - \mathcal{Y}^A Y_A^\dagger\biggr), \hspace{0.5cm}
    \hat{G} = \eth_\mu \hat{A}^\mu + i {\xi} \biggl( Y^A \mathcal{Y}_A^\dagger - \mathcal{Y}^A Y_A^\dagger\biggr),
\end{equation}
with $\eth_\mu$ the covariant derivative with respect to the metric, and $\xi$ a gauge-fixing parameter of dimension one. A similar type of gauge fixing was done in~\cite{Bianchi:2015tba} which treated a case with constant (i.e.\ not space-time dependent)
classical fields. The gauge fixing here is designed to remove terms with derivatives and at the same time avoid complications from 
novel  terms quadratic in $Y$'s. After gauge fixing our Lagrangian takes the form up to ghost terms
\begin{equation}
    \begin{split}
     \mathcal{L}_{ABJM+gf}& = \frac{k}{4 \pi}\Tr \biggl( -\frac{\varepsilon^{\mu\nu\lambda}}{2} \left(A^+_\mu \partial_\nu A^-_\lambda + A_\mu^- \partial_\nu A^+_\lambda \right) + \frac{1}{\xi}\eth_\mu {A^-}^\mu \eth_\nu {A^+}^\nu  + \frac{|\beta|^2}{|z|} A^-_\mu {A^-}^\mu \\
    & + \partial_\mu Y^\dagger_A \partial^\mu Y^A 
     + 2 i (\partial^\mu \mathcal{Y}_A^\dagger )A_\mu^-  Y^A - 2 i (\partial^\mu \mathcal{Y}^A) A_\mu^- Y^\dagger_A    \biggr), 
     \nonumber \end{split}
\end{equation}
or in matrix form exposing more clearly the coupling between fields
\begin{equation} 
    \mathcal{L}_{ABJM+gf} = \frac{k}{4 \pi}\Tr
    \begin{pmatrix}
        A_\mu^- & A^+_\mu  & Y^\dagger_A
    \end{pmatrix} M
    \begin{pmatrix}
        A^-_\lambda \\
        A^+_\lambda \\
        Y^A
    \end{pmatrix},
\end{equation}
where
\begin{equation}
   M=\
    \begin{pmatrix}
        \frac{|\beta|^2}{|z|} \eta^{\mu\lambda} & -\frac{\varepsilon^{\mu\nu\lambda}}{2} \partial_\nu - \frac{1}{2\xi} \eth^\mu \eth^\lambda  & 2i \: \partial^\mu \mathcal{Y}^\dagger_A\\
        - \frac{\varepsilon^{\mu\nu\lambda}}{2}\partial_\nu  - \frac{1}{2\xi} \eth^\mu \eth^\lambda& 0 & 0  \\
        -2i \: \partial^\lambda \mathcal{Y}^A & 0 &  - \Box
    \end{pmatrix}.
\end{equation}
with $\Box = \eth^\nu \eth_\nu$. 
To determine the propagators we are thus faced with a set of differential equations which can compactly be written
as 
\begin{equation}
    \frac{k}{4\pi}M_x G(x,y) = \mathds 1 \delta(x-y).
 \end{equation}
We notice that the parameter $\alpha$ of the background does not appear in the quadratic part of the expanded action. It could only appear either via the covariant derivative acting on a scalar or via the cubic Chern-Simons terms for the gauge fields.\label{alphadependence}. The form of the covariant derivative
for adjoint scalars, cf.\ eqn. (\ref{adjoint}) in combination with the fact that the background
gauge field is proportional to the unit matrix rule out the first possibility. The second possibility is ruled out by the combination of the symmetry properties of the cubic Chern-Simons terms and again the fact that the background gauge field is proportional to the unit matrix, cf.\ page~\pageref{symmetry}. It is easy to check that
$\alpha$ does also not appear, when the action is expanded beyond quadratic order.

We can write the mixing matrix as a sum of the free part and a perturbation proportional to the symmetry breaking parameters $\beta_i$ or $|\beta|^2$. 
 \begin{equation}
    M = M_0 + V,
\end{equation}
where $M_0$ and $V$ are given by
\begin{equation}
    M_0 = \begin{pmatrix}
        0 & {}_x\mathcal{K}^{\mu\lambda}  & 0 \\
        {}_x \mathcal{K}^{\mu \lambda} & 0  & 0\\
        0 & 0 & - \Box \\
    \end{pmatrix}, \qquad 
    V = \begin{pmatrix}
        \frac{|\beta|^2}{|z|} \eta^{\mu \lambda} & 0  & 2i \partial^\mu \mathcal{Y}^\dagger_A \\
        0 & 0 & \hspace{0.3cm}  0 \\
        -2i \partial^\lambda \mathcal{Y}^A & 0&\hspace{0.3cm}  0
    \end{pmatrix}.
\end{equation}
Let us denote the inverse of $M_0$  by $G_0$. It takes the form
\begin{equation}
    G_0(x,y) = \begin{pmatrix}
        0 & G_{\lambda \rho}^{-+}(x,y)  & 0 \\
        G_{\lambda \rho}^{+-}(x,y) & 0  & 0\\
        0 & 0  & G^{Y^\dagger Y}(x,y)
    \end{pmatrix}.
\end{equation}
From $G_0$ we can determine the inverse of the operator $M$
using the Neumann series
\begin{equation}
    M^{-1} = G = G_0 - G_0 V G_0 + G_0 V G_0 V G_0 - ....
\end{equation}
 By direct multiplication, one finds that:
\begin{equation}
    G_0 V G_0 V G_0 V G_0= 0,
\end{equation}
which means that the series truncates after the second correction. This is a generic property of the matrices of the form
\begin{equation}
    G_0 \sim \begin{pmatrix}
      & \:\:\:a  & \\
     a & \:\:\: & \\
     & &\:\:\: b  \\ 
    \end{pmatrix}, \qquad 
    V \sim \begin{pmatrix}
        c & \:\: \:& \:\:\:d \\
        & & \\
        e & & 
    \end{pmatrix}. 
\end{equation}
For simplicity, let us now restrict ourselves to the 1/2-BPS case where only
one scalar, $Y^1$, gets a non-vanishing vacuum expectation value (VEV), and let us denote the corresponding symmetry breaking parameter as $\beta$, i.e.\  $\mathcal{Y}^1=\beta/\sqrt{z}$, we choose $\beta$ to be real. Performing the Neumann expansion we find
\begin{eqnarray}
    G &=& G_0 -G_0 V G_0 + G_0 V G_0 V G_0 \\
    &= &\begin{pmatrix}
        0 & G_{\lambda \rho}^{-+} (x,y)  & 0 \\
         G^{+-}_{\lambda \rho} (x,y) & \:\:\: \beta^2G^{++}_{\lambda \rho}(x,y) \:\:\: & \beta\, G^{+Y}(x,y) \\
         0 & \beta \,G^{Y^\dagger +}(x,y) & G^{Y^\dagger Y}(x,y)
    \end{pmatrix},   
\end{eqnarray}
where we have explicitly exposed the $\beta$-dependence, and where we have suppressed the index on $Y^1$. We notice that the non-vanishing entries of $G_0$ do not get $\beta$ corrections.
At this point, it is natural to ask what class of functions should be admitted as solutions. As the VEV of $Y^1$ has a monodromy, $\mathcal{Y}^1=\beta/\sqrt{z}$,  it is natural to require that the same holds for the associated
quantum fluctuations. This assignment furthermore has the virtue that it  renders the Lagrangian single valued. It can be seen in the following way: expand the sextic potential for the scalar fields to the fifth order in the perturbed fields (the first order is zero by equations of motion and the third order appears to be zero, as one can check). Then one gets terms of the form $\mathcal{Y}^A Y^\dagger_A Y^B Y^\dagger_B Y^C Y^\dagger_C$. If one doesn't assume the same monodromy for the background and  excited fields, one gets a Lagrangian that has a non-trivial transformation when circling around the defect.

To determine the propagator for the scalar $Y^1$ we thus
have to solve the equation
\begin{equation}
    - \Box G^{Y Y^\dagger} (x,y) = \frac{4 \pi}{k} \delta(x-y), 
\end{equation}
with the condition that
\begin{equation}
G^{YY^\dagger}(r,\theta+2\pi,\tau;r',\theta',\tau')
=
-
G^{Y Y^\dagger}(r,\theta,\tau;r',\theta',\tau'),
\end{equation}
where we have now introduced cylindrical coordinates. The general solution can be written in terms of Bessel functions
and is given in appendix~\ref{tadpole}. As the background gauge field does not have a square root branch cut in $z$
we shall assume
the same property of its quantum fluctuations. Then the propagators $G^{+-}_{\lambda \rho}(x,y)$ are standard 
Chern-Simons propagators fulfilling (after gauge-fixing)
\begin{equation}
    {}_x\mathcal{K}^{\mu\lambda} G^{-+}_{\lambda \rho} (x,y) =\bigg(-\frac{\varepsilon^{\mu\nu\lambda}}{2} \partial_\nu - \frac{1}{2 \xi} \eth^\mu \eth^\lambda \bigg) G^{-+}_{\lambda \rho}(x,y) = \frac{4 \pi}{k} \delta^\mu_\rho \delta(x-y),
\end{equation}
with the well-known solution
\begin{equation}\label{G-+}
    G^{-+}_{\lambda \rho}(x,y) = -\frac{2}{k} \bigg( \varepsilon_{\lambda \rho \sigma} \partial^\sigma \frac{1}{|x-y|}-\frac{\xi}{2} \eth_\lambda \eth_\rho |x-y| \bigg).
\end{equation}
From the knowledge of the leading order propagators we can determine the $\beta$-dependent ones as follows 
\begin{equation}
\beta \,G^{+Y}(x,y)=-2i \int d^3v \,G^{+-}_{\lambda \rho}(x,v) \cdot \partial^\rho\mathcal{Y}^\dagger(v) \cdot G^{Y^\dagger Y}(v,y),
\end{equation}
with $G^{Y^\dagger+}(x,y)=G^{+Y}(y,x)$ and 
\begin{eqnarray}
\lefteqn{\beta^2    G^{++}_{\lambda \rho}(x,y) = - \int d^3v \, G^{+-}_{\lambda\sigma}(x,v) \,\frac{\beta^2}{r_v} \eta^{\sigma \theta}_v \,G^{-+}_{\theta \rho} (v,y)} \\
&&   \hspace{0.8cm} + 4 \int d^3v \,d^3w\, G^{+-}_{\lambda \rho}(x,v)\, \partial^\rho \mathcal{Y}^\dagger(v) \,G^{Y^\dagger Y}(v,w)
\,\partial^\sigma \mathcal{Y}(w) G^{-+}_{\sigma \gamma} (w,y).
\end{eqnarray}
Let us point out that in this section we have only considered the simplest possible mixing problem where only one complex scalar, $Y^1$, gets a classical value and where this classical value is proportional to the unit matrix. There are two directions along which generalizations are possible and should in principle be pursued. One generalization consists in allowing the classical  field ${\mathcal{Y}^1}$ to have different elements along its diagonal. This will add to the quantum action (space-time dependent) mass like terms for all the  scalars as well as the gauge fields, and it  will introduce (numerous) additional terms involving derivatives.
In appendix~\ref{general case}, we give an example of the additional mass terms generated for the scalars.
Another generalization consists in allowing non-vanishing classical values for more $Y$-fields (which is relevant for the 1/3-BPS and the 
1/6-BPS cases). This will lead to a non-trivial mixing of the different flavours of $Y$'s.  Needless to say that combining both
of these generalizations entails a yet more involved mixing of both colour and flavour indices. 

  We stress that even though propagators may have
a monodromy, correlation functions of gauge invariant local operators remain single valued. In particular, the spectrum of  conformal dimensions of (gauge invariant) bulk operators is unchanged compared to non-defect ABJM theory.  We have 
previously in section~\ref{onepoint}
computed the leading (in $\lambda$) contribution to one-point functions of such operators and we shall now
address the first quantum correction at the planar level. 

In the simple case that we consider, where only one field $Y^1$ has a classical value and a monodromy, only single trace operators built entirely from this field get a non-zero first quantum correction. We have illustrated the computation in figure~\ref{one-loop}.
\\

\begin{figure}[ht]
\centering
\begin{tikzpicture}[
    thick,
    every node/.style={font=\small},
]

%--------------------------------------------------
% A command to draw a cross
%--------------------------------------------------
\newcommand{\mycross}[2]{
    \draw ($(#1)+(-0.08,-0.08)$)--($(#1)+(0.08,0.08)$);
    \draw ($(#1)+(-0.08,0.08)$)--($(#1)+(0.08,-0.08)$);
}

%% Classical picture

\begin{scope}

\node[fill=black,circle,inner sep=2.2pt] (O) at (0,0) {};

\foreach \a in {0,45,...,315}{
    \draw (O)--++(\a:1.3);
    \mycross{$(O)+(\a:1.3)$}{}
}

\node at (0,-2.0) {(a) The classical expectation value.};

\end{scope}

%% lollipop

\begin{scope}[xshift=5cm]

\node[fill=black,circle,inner sep=2.2pt] (O) at (0,0) {};

% six classical insertions
\foreach \a in {0,45,135,180,225,270,315}{
    \draw (O)--++(\a:1.3);
    \mycross{$(O)+(\a:1.3)$}{}
}

% lollipop
\draw (O)--(0,1.0);
\draw (0,1.5) circle (0.5);

\node at (0,-2.0) {(b) Lollipop diagram};

\end{scope}

%% Tadpole

\begin{scope}[xshift=10cm]

\node[fill=black,circle,inner sep=2.2pt] (O) at (0,0) {};

\foreach \a in {0,180,216,252,288,320}{
    \draw (O)--++(\a:1.3);
    \mycross{$(O)+(\a:1.3)$}{}
}

\draw plot[smooth cycle,tension=0.9] coordinates {
    (0,0)
    (-0.5,0.8)
    (0,1.3)
    (0.5,0.8)
};

\node at (0,-2.0) {(c) Tadpole diagram};

\end{scope}

\end{tikzpicture}
\caption{\label{one-loop} Fig.\ 2(a) shows the classical contribution to the one-point function whereas as Fig.\ 2(b) and Fig.\ 2(c) show the two possible one-loop diagrams. The operator itself is represented by the black circle with lines (fields) radiating from it and the insertion of a classical field
is symbolized by a cross.}
\end{figure}
\noindent
Figure~2(a) shows the leading, classical contribution.
 It is obtained simply by replacing
each field in an operator by its classical value which is symbolized by a line ending in
a cross. At the planar one-loop level there are two possible Feynman diagrams to take 
into account, denoted  as the lollipop diagram and the tadpole diagram. The tadpole diagram
is generated by Wick contracting two neighboring scalar fields from the operator. 
The value of this diagram is vanishing  after dimensional regularization (which is appropriate to conserve supersymmetry) unless the scalar has a monodromy. Only the scalar $Y^1$ has a monodromy so $Y^1$ is the only field which contributes in this loop.
 The lollipop diagram involves a  vertex and both scalars, fermions, gauge fields and ghosts can in principle run in the loop. However, this type of loop diagram is known to vanish due to supersymmetry. The vanishing of the diagram was explicitly demonstrated (by a rather involved and lengthy computation) for the case of a 1/2 BPS defect set-up in ${\cal N}=4$ SYM in~\cite{Buhl-Mortensen:2016jqo}.\footnote{It can also be inferred here. The fully expanded action does not involve any $3Y$-vertices. Furthermore, only the gauge field $A^-$ couples to matter, but the propagators $G^{--}=G^{-Y}=G^{Y^\dagger-}=0$.
The fermion propagators are unaffected by the defect, and the relevant Yukawa vertices introduce no non-trivial spinor structure, so the 
contribution of the massless fermions in the lollipop diagram vanishes as well. The ghosts are also massless and do also not give
any contribution to the lollipop diagram.
}

As demonstrated in appendix~\ref{tadpole}, the renormalized value of the $Y^1$-propagator  with coinciding endpoints (needed for the tadpole diagram) takes the value
\begin{equation}\label{GYY}
G^{Y^\dagger Y}_{ren}(r)=-\frac{1}{k\pi r},
\end{equation}
where we have suppressed the matrix indices. We note that~(\ref{GYY}) 
has the scaling expected for a conformal field theory with a defect. The value of the
tadpole diagram in the planar limit is then a sum over all possible ways of contracting two
neighboring fields inside an operator of length $L$ using the propagator above 
times 
a factor of $N$ coming from the contraction of the matrix indices that we so far suppressed.

 This means that for any one-point
function of an operator built from scalars we can write
\begin{equation}\label{oneloop}
    \langle \mathcal{O}_L(z) \rangle = \langle \mathcal{O}_L(z) \rangle_{\text{classical}}  \bigg( 1 - \frac{2L \lambda}{\pi |\beta|^2} + ... \bigg).
\end{equation}
for the 1/2 BPS monodromy defect defined by eqn.\ (\ref{1/2BPS}) where each classical field is proportional to
 a $N\times N$ unit matrix.
In particular, we do not need to worry about possible corrections to the state for the non-protected operators as such corrections 
only appear at one higher loop order. 
We notice that the correction does not depend on the parameter $\alpha$ of the background,
cf.\ eqn.~(\ref{1/2BPS}). As explained earlier the parameter $\alpha$ never appears in the semi-classical
expansion around the particular  monodromy defect considered, cf.\ page~\pageref{alphadependence}.  This is a special feature of this background owing to the fact that
 the classical fields are proportional to the unit matrix.
Furthermore, we note that the effective loop expansion parameter
in the monodromy background is $\lambda/|\beta|^2$ and therefore we expect the 
semi-classical regime for this background to be given by $\lambda\ll |\beta|^2$. 

\section{A possible extension of the overlap formula\label{extension}}

It has been observed that the general form of the overlap formulas for 
integrable rational bosonic spin chains~\cite{Gombor:2024iix,Gombor:2025wvu} has a natural extension to super spin chains~\cite{Kristjansen:2020vbe,Kristjansen:2021abc,Gombor:2024api}. In analogy with the bosonic case, an eigenstate of a super spin chain is described in terms of a collection of Bethe roots, one type for each node of the Dynkin diagram characterizing the underlying super Lie algebra, and
each node, $a$, carries an associated Baxter polynomial, $Q_a(u)= \prod_{j=1}^{r_a}\left(u-u_j^{(a)}\right)$ corresponding to that eigenstate. The overlap between the normalized Bethe eigenstate and an integrable matrix product
state is then a sum of terms of the type 
\begin{equation}\label{general_form}
\sqrt{\prod_a \frac{\prod_k Q_a(\frac{i\alpha_k}{2})}{\prod_j Q_a(\frac{i\beta_j}{2})}
\, S\det G},
\end{equation}
where the constants $\alpha_k$ and $\beta_j$ have to be determined by means of the KT relation
or otherwise. 
In the case of a matrix product state of bond dimension one, there is only one term in the sum.

An important difference to the bosonic case is that super Lie algebras can be described
in terms of several different Dynkin diagrams. Thus, Bethe eigenstates and their overlap with
matrix product states can be described in different parameterizations. One can translate from
one parametrization to the other, i.e.\ from one Dynkin diagram to another, by means
of fermionic dualities~\cite{Kristjansen:2020vbe}. Under  such a duality after
a specific fermionic node the neighbouring nodes  change their nature from fermionic to bosonic or vice versa. Importantly, it has been demonstrated that
the super determinant of the Gaudin matrix transforms covariantly under fermionic dualities, meaning that a fermionic duality acting on the node $a$ changes the superdeterminant of
the Gaudin matrix in the following way~\cite{Kristjansen:2020vbe}
\begin{equation}\label{duality-overlap}
 \prod_{b:~M_{ab}\neq 0}^{}Q_b(i/2)\,\, \frac{\mathop{\mathrm{Sdet}}\widetilde{G}}{\widetilde{Q}_a(0)}=Q_a(0)\,\mathop{\mathrm{Sdet}}G,
\end{equation}
where the product on the left-hand side is over all nodes adjacent to $a$ and where tilded and
untilded quantities refer to quantities in the two different parametrizations. This means
that the overlap formula retains its form~(\ref{general_form}) only if a fermionic node carries a single $Q$-function with the argument zero. This fact can in some cases be used to predict the overlap formula corresponding to a super Dynkin diagram with a specific grading just from its form on a bosonic sub-diagram, assuming that the overlap formula keeps its factorized form in all gradings.\footnote{It has been pointed out in~\cite{Gombor:2024api} that this assumptions can some times fail.} This type of argument correctly predicts the full overlap formula in the case of the domain wall defect in ${\cal N}=4$ SYM~\cite{Kristjansen:2020vbe}. Furthermore, the strategy leads to a unique prediction for the overlap formula for the ABJM domain wall defect~\cite{Kristjansen:2021abc} and for the 't Hooft line in ${\cal N}=4$ SYM~\cite{Kristjansen:2023ysz,Gombor:2024api}.
In the case of the Gukov-Witten surface defects which could be proven in general to be non-integrable the argument does not allow any conclusion.~\cite{Chalabi:2025nbg}. 

\begin{figure}
 \begin{center}
 \includegraphics[width=12.cm] {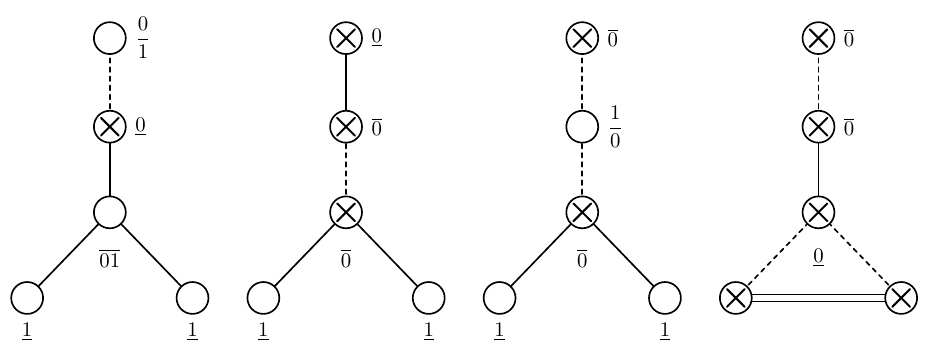}
 \end{center}
\caption{\label{OspDynkin}Four possible Dynkin diagrams of $\mathfrak{osp}(4|6)$ connected by a chain of fermionic dualities and the prediction for the associated overlap formulas. The numbers below or besides the nodes are the arguments of the $Q$-functions entering the overlap formula, and the lines
above or below the numbers indicate whether a $Q$-function appears in the numerator or the denominator}
\end{figure}

Let us apply the argument to our overlap formula~(\ref{overlap}) which we can also write
in the more symmetrical form
\begin{equation}
 \frac{\langle \mathrm{MPS} \vert \mathbf{u}\rangle}
        {\sqrt{\langle \mathbf{u}\vert\mathbf{u}\rangle}}\, \sim \,
        \sqrt{
        \frac{Q_1\left(\frac{i}{2}\right)Q_3\left(\frac{i}{2}\right)}
{Q_2(0)Q_2\left(\frac{i}{2}\right)}S\det G}.
\end{equation}
Four different Dynkin diagrams of $\mathfrak{osp}(4|6)$ connected by fermionic dualities are shown in fig.~\ref{OspDynkin}. To move from the first to the second diagram we dualize after the single fermionic node appearing there, to move from the second to the third we perform a dualization 
after the upper fermionic node and finally we move from the third to the fourth by dualizing
after the lower fermionic node. 
Our known overlap formula is valid for the bosonic sub-diagram with three nodes which appears in the lower part of the first diagram. We indicate by numbers next to or below the nodes of a given diagram the arguments (in units of $\frac{i}{2}$) of the associated $Q$-functions which appear (under the square root) in the overlap formula. The lines above or below the numbers
indicate whether a $Q$-function appears in the numerator or in the denominator.
In order for the overlap to stay on the covariant form when a dualization after a fermionic node is performed the original fermionic node can only have a $Q$-function with
argument zero in either the numerator or the denominator. If more $Q$-functions are present the transformation rule~(\ref{duality-overlap}) will not lead to a complete reparametrization of the overlap and a much more complicated overlap formula can be 
expected.
In our case there is only one  way to associate $Q$-functions to the upper
two nodes in the first Dynkin diagram which ensures that the overlap formula stays on the
covariant form and that fermionic dualization is possible at any fermionic node in each step.

The fact that it is possible to assign
a set of Q-functions to all  Dynkin diagrams above is very encouraging. In particular, it is important that the assignment works for the third Dynkin
diagram which is the one that allows one to solve the spectral problem to all loop orders~\cite{Gromov:2008qe,Ahn:2008aa}.
This raises a hope for higher loop integrability of the one-point functions.
Our prediction for the overlap formula corresponding to the third Dynkin diagram is more precisely
\begin{equation}
 \frac{\langle \mathrm{MPS} \vert \mathbf{u}\rangle}
        {\sqrt{\langle \mathbf{u}\vert\mathbf{u}\rangle}}\, \sim \,
        \sqrt{
        \frac{Q_1\left(\frac{i}{2}\right)Q_3\left(\frac{i}{2}\right)Q_4\left(\frac{i}{2}\right)}
{Q_2(0)\,Q_4\left(0\right)Q_5\left(0\right)}S\det G},
\end{equation}
where the numbering of nodes is as in figure~\ref{SU4Dynkin} supplemented by
numbers $5$ and $4$ for the top node and the one just beneath.

\section{Conclusion and Outlook\label{conclusion}}

We have derived a closed form expression for the leading order contribution to one-point functions of non-protected scalar operators in ABJM theory in the presence of supersymmetric monodromy defects. Furthermore, we have explained how to perform the perturbative quantization around these defects and carried out the procedure
for the simplest case of background fields proportional to the unit matrix.
While higher order perturbative computations for monodromy backgrounds are  in principle feasible it requires a strong motivation to proceed with the rather extensive book-keeping involved for the general setting. It would be more satisfactory if the higher
loop contributions to the one-point functions could be found by integrability bootstrap
as it was the case for the domain wall defect in ${\cal N}=4$ SYM theory~\cite{Gombor:2020kgu,Komatsu:2020sup,Gombor:2020auk}. 
A first indication of whether this could be possible can be obtained by investigating whether our overlap formula has a natural extension 
through fermionic duality relations~\cite{Kristjansen:2020vbe} to the particular $\mathfrak{osp}(4|6)$ Dynkin diagram
which was used to solve the spectral problem of ABJM theory at all loop orders~\cite{Gromov:2008qe}.
We found in Section~\ref{extension} that this is indeed the case.

Another possible direction of investigation would be to study 
 the interplay between the supersymmetric monodromy defects and supersymmetric Wilson loops. Like the monodromy defects, the Wilson loops come  with different amounts
of supersymmetry, namely 1/2 BPS~\cite{Drukker:2009hy}, 1/3 BPS~\cite{Drukker:2022txy} and 1/6 BPS~\cite{Drukker:2008zx,Chen:2008bp,Rey:2008bh}, see also~\cite{Ouyang:2015iza,Ouyang:2015bmy,Mauri:2018fsf}, and a lot of expertise on Wilson loops in ABJM theory has been gathered in recent years~\cite{Drukker:2019bev}.

Needless to say that it would be interesting if our overlap formula could be compared to one-point functions computed in the string theory interpretation of ABJM theory. The monodromy defects have a string theory description as a $M2$-brane ending on the boundary of an $AdS_4$-space~\cite{Drukker:2008jm}. While it is possible to compute one-point functions  of chiral primaries from the string theory perspective~\cite{Drukker:2008jm} devising a  strategy for 
non-protected operators is  challenging. Let us note that even 
for protected operators there appears to be a non-trivial difference in the functional dependence of
one-point functions on $\beta$ and $\lambda$ and no signs of numerical factors agreeing either. Likewise for the domain wall defect in ABJM theory, there was no simple relation between the leading 
order results for one-point functions of protected operators at weak and strong coupling~\cite{Kristjansen:2021abc}. It would be interesting if
these two types of supersymmetric defects in ABJM theory could be handled by supersymmetric localization by an adaptation of the techniques developed in~\cite{Kapustin:2009kz,Marino:2011eh,Drukker:2010nc}.

\acknowledgments

The authors were supported by 
Villum Fonden via the Villum Investigator grant 73742.
We thank Jan Ambj\o rn, Nadav Drukker, Peter Orland, Chenliang Su and Konstantin Zarembo for useful discussions.

\appendix

\section{\label{check} Verification of the overlap formula for short operators.}

In this appendix we compute the one-point function of short operators directly in the field theory language and verify
that the result is reproduced by the formula~(\ref{O(x)}) with the overlap given by eqn.~(\ref{overlap}). We will make use
of the short conformal operators considered in~\cite{Minahan:2008hf} for which the Bethe roots are known analytically.
 Let us start with the following field theory normalized operator of length $L=2$ 
\begin{equation}
    \mathcal{O}_2 = \frac{1}{4\sqrt{5}\lambda^2} \Tr(Y^AY_A^\dagger Y^B Y_B^\dagger + Y^A Y^\dagger_B Y^B Y_A^\dagger).
\end{equation}
At the leading order the one-point function is simply found by replacing the fields by their classical values. Clearly,
the $z-$dependence will be the same for all the symmetry breaking cases, namely $1/|z|^2$. 
The one-point function is thus  equal to 
\begin{equation}
    \langle \mathcal{O}_2(x) \rangle = \frac{ 1}{2\sqrt{5}\lambda^2} \frac{1}{|z|^2}  \sum_{k=1}^M N_k  |\beta^{(k)}|^4 . \label{ex1}
\end{equation}
The operator above corresponds to a spin chain eigenstate with  Bethe roots taking the values
\begin{equation}
    \begin{split}
        & \{ u^{(1)} \} = \left\{ \sqrt{\frac{3}{20}}, -\sqrt{\frac{3}{20}} \right\} = \{-u^{(3)}\}, \hspace{0.5cm}
         \{u^{(2)}\} = \left\{ \sqrt{\frac{1}{5}}, -\sqrt{\frac{1}{5}} \right\}.
    \end{split}
\end{equation}
In particular, this state fulfills the selection rule $r_1=r_2=r_3=L$, which is in accordance with the fact the result~(\ref{ex1}) 
above is
non-vanishing for all three defect set-ups.
 The Gaudin superdeterminant is found to be
\begin{equation}
    \sqrt{S \det G} = \sqrt{\frac{9}{10}}.
\end{equation}
and the necessary product of $\tilde{F}$-functions and vacuum overlap takes the value
\begin{equation}
\langle \mbox{MPS}_j| {\Omega}\rangle \, \prod_{k=1}^{2} \widetilde{\mathcal{F}}^{(1)} (u^{(1)}_k)  \prod_{l=1}^{1} \widetilde{\mathcal{F}}^{(2)} (u_{l}^{(2)}) =\frac{1}{3} |\beta^{(j)}|^4,
\end{equation}
for all three BPS set-ups.
The value of the one-point function as read of from the overlap formula therefore reads
\begin{equation}
    \langle \mathcal{O}_2(x) \rangle = \sum_{j=1}^{d_B} \frac{1}{\sqrt{L} \lambda^L}
    \frac{\bra{\mbox{MPS}_j} \mathbf{u} \rangle}{\sqrt{\bra{\mathbf{u}} \mathbf{u}\rangle}}=
        \frac{ 1}{2\sqrt{5}\lambda^2 |z|^2} \sum_{k=1}^M N_k |\beta^{(k)}|^4, 
\end{equation}
where the first sum is over $d_B=N$ overlaps for matrix product states of bond dimension one. 
We see that the results match perfectly.

Next, let us perform some tests which involve special roots. We start with an operator for which the
corresponding root configuration has an unpaired zero root, namely the following operator of  length $L=3$:
\begin{equation}
    \begin{split}\label{L6operator}
    & \mathcal{O}_3(x) = \frac{1}{ 3 \sqrt{80(46-\sqrt{10})}\lambda^3} ((\sqrt{10}-1) \; \text{Tr} (Y^AY_A^\dagger Y^B Y_B^\dagger Y^C Y_C^\dagger + Y^A Y_C^\dagger Y^B Y_A^\dagger Y^C Y_B^\dagger - \\
&    - Y^A Y_B^\dagger Y^B Y_C^\dagger Y^C Y_A^\dagger) + 9 \; \text{Tr} (Y^A Y_A^\dagger Y^B Y_C^\dagger Y^C Y_B^\dagger) ).
    \end{split}
\end{equation}
By substituting the fields by the solutions of the BPS equations we get
\begin{equation}
    \langle \mathcal{O}_3(x) \rangle = \frac{1}{\lambda^3}\frac{\sqrt{10}+8}{3\sqrt{80(46-\sqrt{10})}} \frac{1}{|z|^3} \sum_{k=1}^{M}N_k |\beta^{(k)}|^6.
\end{equation}
The operator~(\ref{L6operator}) is characterized by the Bethe roots:
\begin{equation}
    \{ u^{(1)}\} = \{ \alpha, -\alpha, 0 \} = \{ -u^{(3)}\}, \;\;\;\; \{u^{(2)}\} = \left\{ \frac{2}{\sqrt{3}}\alpha, -\frac{2}{\sqrt{3}}\alpha, 0\right\}, \;\;\;\; \alpha^2 = \sqrt{\frac{2}{5}}-\frac{1}{4}.
\end{equation}
The unpaired zero-root at the middle (chiral) node requires special treatment~\cite{deLeeuw:2016umh,DeLeeuw:2018cal,Kristjansen:2021xno}. First, one should leave out the zero root in
$\tilde{\cal F}^{(2)}$. Secondly, one should compute
the super determinant of the Gaudin matrix following the special recipe in~\cite{deLeeuw:2016umh,DeLeeuw:2018cal,Kristjansen:2021xno}
which gives
\begin{equation}
  S  \det G = \frac{2}{65}(214+25\sqrt{10}).
\end{equation}
Evaluating the one-point function using the overlap formula (\ref{O(x)}) we then get (leaving out the phase $(-1)$)
\begin{equation}
\langle {\cal O}_3(x)\rangle=  \sum_{j=1}^{d_B} \frac{1}{\lambda^L \sqrt{L}}   \frac{\bra{\mbox{MPS}_j} \mathbf{u} \rangle}{\sqrt{\bra{\mathbf{u}}\mathbf{u} \rangle}} = \frac{1}{4 \lambda^3} \sqrt{\frac{1}{65}(22+5 \sqrt{10})}\frac{1}{|z|^3} \sum_{k=1}^M N_k |\beta^{(k)}|^6,
\end{equation}
which agrees with the field theory result.

Finally, let
us consider an example where singular roots appear. At $L=2$ there is
a conformal operator (orthogonal to the one considered above) which takes the form
\begin{equation}
    \mathcal{O}^{sp}_2(x) = \bigg( \frac{2 \pi}{\lambda} \bigg)^2 \frac{1}{4\sqrt{3}} \Tr(Y^A Y^\dagger_A Y^B Y^\dagger_B - Y^A Y^\dagger_B Y^B Y^\dagger_A),
\end{equation}
and which corresponds to a root configuration with singular roots  on the outer nodes~\cite{Minahan:2008hf}. The one-point function of this operator computed by inserting the classical fields clearly
vanishes.
\begin{equation}
\langle {\cal O}_2^{sp}(x)\rangle=0.
\end{equation}
As already mentioned singular roots require a regularization already at the level of the Bethe equations. In order to apply the overlap formula to a regularized version of the root configuration we will need a 
regularization procedure which conserves the achiral root structure which
is not the case for the one considered in~\cite{Minahan:2008hf}. We shall follow
the strategy of~\cite{Kristjansen:2021xno} and add impurities of alternating signs along the sub-chain consisting of the  fundamental sites as well as along the sub-chain consisting of the anti-fundamental sites, so that the left hand side of the
Bethe equations~(\ref{BEs}) for the outer nod
es is changed to
\begin{equation}
\frac{(u_k^{(\alpha)}-i/2)^2 - \theta^2}{(u_k^{(\alpha)}+i/2)^2 - \theta^2},
\hspace{0.5cm} \alpha=1,3, \hspace{0.5cm} k=1,2.
\end{equation}
With this regularization one finds for the roots to order $\theta^2$
\begin{equation}
    \begin{split}
    & \{ u^{(1)} \} = \{ -u^{(3)} \} = \bigg\{ \frac{i}{2}(1-\theta^2), -\frac{i}{2}(1-\theta^2) \bigg\}, \\
    & \{ u^{(2)} \} = \bigg \{ \frac{i}{\sqrt{3}}(1-\theta^2), -\frac{i}{\sqrt{3}}(1-\theta^2) \bigg\}.
    \end{split}
\end{equation}
We can thus compute the components of the Gaudin determinant which are needed to form the superdeterminant using the recipe of~\cite{Kristjansen:2021abc} and find
\begin{equation}
    \det G^+ = -\frac{36}{\theta^2} + \underline{\underline{ \mathcal{O}}}(\theta^0), \qquad \det G^- = -\frac{216}{\theta^2}+ \underline{\underline{ \mathcal{O}}} (\theta^{0}).
\end{equation}
We see that the individual components of the Gaudin determinat are divergent as $\theta\rightarrow 0$ but the superdeterminant 
$S\det G= \det G_+/\det G_-$
is not 
\begin{equation}
    \lim_{\theta \rightarrow 0}\,\, \frac{\det G^+}{\det G^-} = \frac{1}{6}. 
\end{equation}
In the overlap formula the Gaudin superdeterminant is multiplied by 
$Q_1(i/2)Q_3(i/2)$
which vanishes as $\theta\rightarrow 0$ and therefore the overlap fomula tells us that $\langle{\cal O}_2^{sp}(x)\rangle=0$
in accordance with the field theory result.

\section{The tadpole in the presence of the monodromy \label{tadpole}}

Here we will compute the regularized tadpole diagram in the presence of the monodromy defect. We thus wish to solve
the differential equation for the propagator
\begin{equation}
- \Box
G(r,\theta,\tau;r',\theta',\tau')
=
 \frac{4 \pi}{k} \frac{1}{r}\,
\delta(r-r')\,
\delta(\theta-\theta')\,
\delta(\tau-\tau'),
\end{equation}
with the boundary condition 
\begin{equation}
G(r,\theta+2\pi,\tau;r',\theta',\tau')
=
-
G(r,\theta,\tau;r',\theta',\tau')
\end{equation}
The general solution can be written in terms of Bessel functions in the following way
\begin{align}
G_{\frac12}
\bigl(r,\theta,\tau;r',\theta',\tau'\bigr)
&=
\frac{2}{k}
\sum_{m\in\mathbb{Z}+\frac12}
e^{im(\theta-\theta')}
\int_{-\infty}^{\infty}
\frac{d\omega}{2\pi}\,
e^{i\omega(\tau-\tau')}
\int_{0}^{\infty}
dk\,k\,
\frac{
J_{|m|}(kr)J_{|m|}(kr')
}{
k^2+\omega^2
}
\nonumber\\[4pt]
&=
\frac{1}{k}
\sum_{m\in\mathbb{Z}+\frac12}
e^{im(\theta-\theta')}
\int_{0}^{\infty}
dk\,
e^{-k|\tau-\tau'|}
J_{|m|}(kr)J_{|m|}(kr').
\end{align}
The difference to the standard case without the monodromy defect lies only in the sum over half-integer modes instead of
integer modes.  We will regularize the coincident propagator by subtracting the standard propagator before taking the coincidence limit, i.e.\
\begin{align}
G_{\mathrm{ren}}(r)
&=
\lim_{x'\to x}
\left[
G_{\frac12}(x,x')-G_0(x,x')
\right]
\nonumber\\[4pt]
&=
\frac{1}{kr}
\int_0^\infty dq
\left[
2\sum_{n=0}^{\infty}J_{n+\frac12}(q)^2
-
J_0(q)^2
-
2\sum_{n=1}^{\infty}J_n(q)^2
\right] \\
&= \frac{1}{kr}
\int_0^\infty dq
\left[
2\sum_{n=0}^{\infty}J_{n+\frac12}(q)^2
-1
\right]
\\
&\equiv \frac{1}{kr}
\int_0^\infty dq \, F(q).
\end{align}
Differentiating we find
\begin{equation}
F'(q) =\frac{2}{\pi q} \sin (2q), \hspace{0.5cm} F(0)=-1.
\end{equation}
More precisely, the expression for the derivative is found by differentiating the sum over Bessel functions and subsequently applying  the standard two-step recursion relation for the derivative of a Bessel function which results in a single  telescoping sum. Thus
\begin{equation}
F(q)= -1+ \frac{2}{\pi} \int_0^q \frac{\sin (2t)}{t} = \frac{2}{\pi}\,\mbox{Si}(2q)-1.
\end{equation}
Let us introduce an upper cut-off $R$ on the $q$ integral. Then we get
\begin{align}
I(R)
&=
\int_0^R dq\,F(q)
\nonumber\\
&=
\left[qF(q)\right]_0^R
-
\int_0^R dq\,qF'(q)
\nonumber\\
&=
R\,F(R)
-
\frac{2}{\pi}\int_0^R dq\,\sin(2q)
\nonumber\\
&=
R\,F(R)
-
\frac{1}{\pi}\bigl(1-\cos(2R)\bigr),
\end{align}
The large-R behaviour of $\mbox{Si}(2R)$ is well-known

\begin{equation}
\operatorname{Si}(2R)
=
\frac{\pi}{2}
-
\frac{\cos(2R)}{2R}
-
\frac{\sin(2R)}{(2R)^2}
+
\mathcal{O}\!\left(\frac{1}{R^3}\right),
\end{equation}
which implies that
\begin{equation}
I(R) = -\frac{1}{\pi}+
\mathcal{O}\!\left(\frac{1}{R}\right),
\end{equation}
and thus
\begin{equation}
G_{ren}(r)=-\frac{1}{\pi k r}.
\end{equation}

\section{Quantization in the general case \label{general case}}

While the perturbative quantization around a more general monodromy defect, where
the VEVs are not proportional to the unit matrix, in principle poses no novel conceptual barriers, its implementation introduces a proliferation of mixing terms which makes the necessary book-keeping immense. Let us illustrate this
by a brief consideration of the simplest such example, a 1/2-BPS configuration where the only non-vanishing classical scalar field takes the form
\begin{equation}
    \mathcal{Y}^1 = \frac{1}{\sqrt{z}} \begin{pmatrix}
        \beta^{(1)} \mathds{1}_{N_1 \times N_1} & 0 \\
        0 & \beta^{(2)} \mathds{1}_{N_2 \times N_2}
    \end{pmatrix}
\end{equation}
and similarly 
\begin{equation}
    \mathcal{A}^+_z = - \frac{i}{2kz} \begin{pmatrix}
        \alpha_1 & {} \\
        {} & \alpha_2
    \end{pmatrix}, \hspace{0.5cm}
    \mathcal{A}^+_{\bar{z}} =  \frac{i}{2k \bar{z}} \begin{pmatrix}
        \alpha_1 & {} \\
        {} & \alpha_2
    \end{pmatrix},\hspace{0.5cm}
    \mathcal{A}_t^+ = - \frac{1}{|z|} \begin{pmatrix}
        |\beta_1|^2 & {} \\
        {} & |\beta_2|^2
    \end{pmatrix},
\end{equation}
where for simplicity we have left out the unit-matrices. To quantize around this
type of classical configuration we will have to decompose our fields in block form as
\begin{equation}
    Y^A = \begin{pmatrix}
    Y^A_{\nwarrow} & Y^A_{\nearrow} \\
    Y^A_{\swarrow} & Y^A_{\searrow}
    \end{pmatrix}, 
    \qquad \qquad 
    Y^\dagger_A = \begin{pmatrix}
    Y^\dagger_{A \nwarrow} & Y^\dagger_{A\swarrow} \\
    Y^\dagger_{A\nearrow} & Y^\dagger_{A \searrow}
    \end{pmatrix},
\end{equation}
as the individual blocks have different mixing structure. 
The most simple new
type of terms one obtains with VEVs of the type above are (space-time dependent)
mass-like terms for both scalars and gauge fields. E.g.\ for the scalars these
terms read
\begin{equation}
    \begin{split}
    & S_{mass}^{scalar} = \frac{k}{4\pi} \int d^3 x \Bigg[- \frac{(\beta_1^2 -\beta_2^2)^2}{4 |z|^2} \bigg( \Tr( Y^A_\nearrow Y^\dagger_{A \nearrow} ) +\Tr(Y^A_{\swarrow} Y^\dagger_{A\swarrow})\bigg) +\\
    & + \frac{(\beta_1^2 - \beta_2^2)^2}{4 |z|^2} \bigg( \Tr(Y^1_\nearrow Y^\dagger_{1 \nearrow})+\Tr(Y^1_\swarrow Y^\dagger_{1 \swarrow})\bigg) + \\
    & + \frac{1}{|z|^2} \bigg( \Tr(Y^A_\nearrow Y^\dagger_{A\nearrow})+\Tr(Y^A_\swarrow Y^\dagger_{A\swarrow}) \bigg) \bigg\{ - (\beta_1^2-\beta_2^2)^2 - \frac{(\alpha_1-\alpha_2)^2}{k^2}\bigg\}  \bigg],
    \end{split}
\end{equation}
where the sum over $A$ is understood. Here the first two lines originate from the sextic scalar potential and the third one from the covariant derivatives acting on the scalars. Such mass terms are well-known from other defect set-ups, see
e.g.~\cite{Buhl-Mortensen:2016pxs}. 
For a scalar field that does not mix with the gauge fields, the only effect of such terms is a shift in the indices of the Bessel functions appearing in the propagators (see Appendix~\ref{tadpole}). The propagation of such a field can therefore be interpreted as that of a particle with an effective mass, determined by the $\alpha$'s and $\beta$'s, moving in an effective AdS$_3$  geometry with the defect located at its boundary.
In the present case, however, expanding the covariant derivatives in the original action generates numerous coupling terms between all scalar fields and the gauge fields, including derivative interactions. We shall refrain from presenting the fully expanded action.

\bibliographystyle{JHEP}

% Bibliography

% [\alpha] Recommended: using JHEP.bst file

\bibliography{Monodromy.bib}

@article{DeLeeuw:2018cal,
    author = "De Leeuw, Marius and Kristjansen, Charlotte and Linardopoulos, Georgios",
    title = "{Scalar one-point functions and matrix product states of AdS/dCFT}",
    eprint = "1802.01598",
    archivePrefix = "arXiv",
    primaryClass = "hep-th",
    reportNumber = "TCDMATH-18-04",
    doi = "10.1016/j.physletb.2018.03.083",
    journal = "Phys. Lett. B",
    volume = "781",
    pages = "238--243",
    year = "2018"
}

@article{Brockmann_2014,
   title={A Gaudin-like determinant for overlaps of Néel and XXZ Bethe states},
   volume={47},
   ISSN={1751-8121},
   url={http://dx.doi.org/10.1088/1751-8113/47/14/145003},
   DOI={10.1088/1751-8113/47/14/145003},
   number={14},
   journal={Journal of Physics A: Mathematical and Theoretical},
   publisher={IOP Publishing},
   author={Brockmann, M and De Nardis, J and Wouters, B and Caux, J-S},
   year={2014},
   month=Mar, pages={145003} }

@article{Liu:2026fxi,
    author = "Liu, Yang and Bai, Nan and Shao, Mao-Zhong and Wu, Jun-Bao",
    title = "{Chiral Integrable Boundary States of ABJM Spin Chain from Reflection Equations}",
    eprint = "2602.01697",
    archivePrefix = "arXiv",
    primaryClass = "hep-th",
    reportNumber = "USTC-ICTS/PCFT-26-07",
    month = "2",
    year = "2026"
}

@article{Kristjansen:2023ysz,
    author = "Kristjansen, Charlotte and Zarembo, Konstantin",
    title = "{{\textquoteright}t Hooft loops and integrability}",
    eprint = "2305.03649",
    archivePrefix = "arXiv",
    primaryClass = "hep-th",
    doi = "10.1007/JHEP08(2023)184",
    journal = "JHEP",
    volume = "08",
    pages = "184",
    year = "2023"
}

@article{Mauri:2018fsf,
    author = "Mauri, Andrea and Ouyang, Hao and Penati, Silvia and Wu, Jun-Bao and Zhang, Jiaju",
    title = "{BPS Wilson loops in $ \mathcal{N} $ {\ensuremath{\geq}} 2 superconformal Chern-Simons-matter theories}",
    eprint = "1808.01397",
    archivePrefix = "arXiv",
    primaryClass = "hep-th",
    reportNumber = "CJQS-2018-013",
    doi = "10.1007/JHEP11(2018)145",
    journal = "JHEP",
    volume = "11",
    pages = "145",
    year = "2018"
}

@article{Ouyang:2015bmy,
    author = "Ouyang, Hao and Wu, Jun-Bao and Zhang, Jia-ju",
    title = "{Construction and classification of novel BPS Wilson loops in quiver Chern{\textendash}Simons-matter theories}",
    eprint = "1511.02967",
    archivePrefix = "arXiv",
    primaryClass = "hep-th",
    doi = "10.1016/j.nuclphysb.2016.07.018",
    journal = "Nucl. Phys. B",
    volume = "910",
    pages = "496--527",
    year = "2016"
}

@article{Ouyang:2015iza,
    author = "Ouyang, Hao and Wu, Jun-Bao and Zhang, Jia-ju",
    title = "{Novel BPS Wilson loops in three-dimensional quiver Chern{\textendash}Simons-matter theories}",
    eprint = "1510.05475",
    archivePrefix = "arXiv",
    primaryClass = "hep-th",
    doi = "10.1016/j.physletb.2015.12.021",
    journal = "Phys. Lett. B",
    volume = "753",
    pages = "215--220",
    year = "2016"
}

@article{Ahn:2008aa,
    author = "Ahn, Changrim and Nepomechie, Rafael I.",
    title = "{N=6 super Chern-Simons theory S-matrix and all-loop Bethe ansatz equations}",
    eprint = "0807.1924",
    archivePrefix = "arXiv",
    primaryClass = "hep-th",
    reportNumber = "UMTG-258",
    doi = "10.1088/1126-6708/2008/09/010",
    journal = "JHEP",
    volume = "09",
    pages = "010",
    year = "2008"
}

@article{Jiang:2023cdm,
    author = "Jiang, Yunfeng and Wu, Jun-Bao and Yang, Peihe",
    title = "{Wilson-loop one-point functions in ABJM theory}",
    eprint = "2306.05773",
    archivePrefix = "arXiv",
    primaryClass = "hep-th",
    reportNumber = "USTC-ICTS/PCFT-23-11",
    doi = "10.1007/JHEP09(2023)047",
    journal = "JHEP",
    volume = "09",
    pages = "047",
    year = "2023"
}

@article{Kristjansen:2020vbe,
    author = {Kristjansen, Charlotte and M{\"u}ller, Dennis and Zarembo, Konstantin},
    title = "{Overlaps and fermionic dualities for integrable super spin chains}",
    eprint = "2011.12192",
    archivePrefix = "arXiv",
    primaryClass = "hep-th",
    reportNumber = "NORDITA 2020-109",
    doi = "10.1007/JHEP03(2021)100",
    journal = "JHEP",
    volume = "03",
    pages = "100",
    year = "2021"
}

@article{Marino:2011eh,
    author = "Marino, Marcos and Putrov, Pavel",
    title = "{ABJM theory as a Fermi gas}",
    eprint = "1110.4066",
    archivePrefix = "arXiv",
    primaryClass = "hep-th",
    doi = "10.1088/1742-5468/2012/03/P03001",
    journal = "J. Stat. Mech.",
    volume = "1203",
    pages = "P03001",
    year = "2012"
}

@article{Drukker:2010nc,
    author = "Drukker, Nadav and Marino, Marcos and Putrov, Pavel",
    title = "{From weak to strong coupling in ABJM theory}",
    eprint = "1007.3837",
    archivePrefix = "arXiv",
    primaryClass = "hep-th",
    reportNumber = "HU-EP-10-39",
    doi = "10.1007/s00220-011-1253-6",
    journal = "Commun. Math. Phys.",
    volume = "306",
    pages = "511--563",
    year = "2011"
}

@article{Kapustin:2009kz,
    author = "Kapustin, Anton and Willett, Brian and Yaakov, Itamar",
    title = "{Exact Results for Wilson Loops in Superconformal Chern-Simons Theories with Matter}",
    eprint = "0909.4559",
    archivePrefix = "arXiv",
    primaryClass = "hep-th",
    reportNumber = "CALT-68-2750",
    doi = "10.1007/JHEP03(2010)089",
    journal = "JHEP",
    volume = "03",
    pages = "089",
    year = "2010"
}

@article{Gromov:2008qe,
    author = "Gromov, Nikolay and Vieira, Pedro",
    title = "{The all loop AdS4/CFT3 Bethe ansatz}",
    eprint = "0807.0777",
    archivePrefix = "arXiv",
    primaryClass = "hep-th",
    reportNumber = "LPTENS-08-39",
    doi = "10.1088/1126-6708/2009/01/016",
    journal = "JHEP",
    volume = "01",
    pages = "016",
    year = "2009"
}

@article{Drukker:2019bev,
    author = "Drukker, Nadav and others",
    title = "{Roadmap on Wilson loops in 3d Chern{\textendash}Simons-matter theories}",
    eprint = "1910.00588",
    archivePrefix = "arXiv",
    primaryClass = "hep-th",
    doi = "10.1088/1751-8121/ab5d50",
    journal = "J. Phys. A",
    volume = "53",
    number = "17",
    pages = "173001",
    year = "2020"
}

@article{Andrei:2018die,
    author = "Andrei, N. and others",
    title = "{Boundary and Defect CFT: Open Problems and Applications}",
    eprint = "1810.05697",
    archivePrefix = "arXiv",
    primaryClass = "hep-th",
    doi = "10.1088/1751-8121/abb0fe",
    journal = "J. Phys. A",
    volume = "53",
    number = "45",
    pages = "453002",
    year = "2020"
}

@article{Dixon:1987qvj,
  author       = {Dixon, Lance J. and Friedan, Daniel and Martinec, Emil J. and Shenker, Stephen H.},
  title        = {The Conformal Field Theory of Orbifolds},
  journal      = {Nucl. Phys. B},
  volume       = {282},
  pages        = {13--73},
  year         = {1987},
  doi          = {10.1016/0550-3213(87)90676-6},
  reportNumber = {EFI-86-41-CHICAGO},
}

@article{Dixon:1985jw,
  author       = {Dixon, Lance J. and Harvey, Jeffrey A. and Vafa, Cumrun and Witten, Edward},
  title        = {Strings on Orbifolds},
  journal      = {Nucl. Phys. B},
  volume       = {261},
  pages        = {678--686},
  year         = {1985},
  doi          = {10.1016/0550-3213(85)90593-0},
  reportNumber = {HUTP-85/A059},
}

@article{Lauria:2020emq,
    author = "Lauria, Edoardo and Liendo, Pedro and Van Rees, Balt C. and Zhao, Xiang",
    title = "{Line and surface defects for the free scalar field}",
    eprint = "2005.02413",
    archivePrefix = "arXiv",
    primaryClass = "hep-th",
    reportNumber = "CPHT-RR003.012021, DESY-20-082",
    doi = "10.1007/JHEP01(2021)060",
    journal = "JHEP",
    volume = "01",
    pages = "060",
    year = "2021"
}

@misc{BianchiTalk,
  author       = {Lorenzo Bianchi},
  title        = {Analytic bootstrap for holographic surface defect},
  howpublished = {Talk presented at the Workshop Defects and Extended Excitations in Quantum Field Theory, Quantum Matter and Statistical Models, Florence, Italy},
  month        = may,
  year         = {2026},
}

@inbook{Gukov:2014gja,
    author = "Gukov, Sergei",
    editor = {Teschner, J{\"o}rg},
    title = "{Surface operators.}",
    booktitle = "{New Dualities of Supersymmetric Gauge Theories}",
    eprint = "1412.7127",
    archivePrefix = "arXiv",
    primaryClass = "hep-th",
    doi = "10.1007/978-3-319-18769-3_8",
    pages = "223--259",
    year = "2016"
}

@article{Bianchi:2015tba,
    author = "Bianchi, Marco S.",
    title = "{A note on scattering amplitudes on the moduli space of ABJM}",
    eprint = "1502.05352",
    archivePrefix = "arXiv",
    primaryClass = "hep-th",
    reportNumber = "QMUL-PH-15-03",
    doi = "10.1007/JHEP06(2015)194",
    journal = "JHEP",
    volume = "06",
    pages = "194",
    year = "2015"
}

@article{Kristjansen:2024zvl,
    author = "Kristjansen, Charlotte and Qian, Xin and Su, Chenliang",
    title = "{The spectrum of defect ABJM theory}",
    eprint = "2412.17479",
    archivePrefix = "arXiv",
    primaryClass = "hep-th",
    doi = "10.1007/JHEP05(2025)207",
    journal = "JHEP",
    volume = "05",
    pages = "207",
    year = "2025"
}

@article{Kristjansen:2021xno,
    author = {Kristjansen, Charlotte and M{\"u}ller, Dennis and Zarembo, Konstantin},
    title = "{Duality relations for overlaps of integrable boundary states in AdS/dCFT}",
    eprint = "2106.08116",
    archivePrefix = "arXiv",
    primaryClass = "hep-th",
    reportNumber = "NORDITA 2021-042",
    doi = "10.1007/JHEP09(2021)004",
    journal = "JHEP",
    volume = "09",
    pages = "004",
    year = "2021"
}

@article{deLeeuw:2016umh,
    author = "de Leeuw, Marius and Kristjansen, Charlotte and Mori, Stefano",
    title = "{AdS/dCFT one-point functions of the SU(3) sector}",
    eprint = "1607.03123",
    archivePrefix = "arXiv",
    primaryClass = "hep-th",
    doi = "10.1016/j.physletb.2016.10.044",
    journal = "Phys. Lett. B",
    volume = "763",
    pages = "197--202",
    year = "2016"
}

@inproceedings{Kristjansen:2024dnm,
    author = "Kristjansen, Charlotte and Zarembo, Konstantin",
    title = "{Integrable Holographic Defect CFTs}",
    booktitle = "{Gravity, Strings and Fields}: {A Conference in Honour of Gordon Semenoff}",
    eprint = "2401.17144",
    archivePrefix = "arXiv",
    primaryClass = "hep-th",
    doi = "10.1007/978-3-031-91266-5_6",
    month = "1",
    year = "2024"
}

@article{DeLeeuw:2019ohp,
    author = "De Leeuw, Marius and Gombor, Tam{\'a}s and Kristjansen, Charlotte and Linardopoulos, Georgios and Pozsgay, Bal{\'a}zs",
    title = "{Spin Chain Overlaps and the Twisted Yangian}",
    eprint = "1912.09338",
    archivePrefix = "arXiv",
    primaryClass = "hep-th",
    doi = "10.1007/JHEP01(2020)176",
    journal = "JHEP",
    volume = "01",
    pages = "176",
    year = "2020"
}

@article{Gombor:2025wvu,
    author = "Gombor, Tamas",
    title = "{Derivations for the MPS overlap formulas of rational spin chains}",
    eprint = "2505.20234",
    archivePrefix = "arXiv",
    primaryClass = "hep-th",
    doi = "10.1007/JHEP10(2025)035",
    journal = "JHEP",
    volume = "10",
    pages = "035",
    year = "2025"
}

@article{Gombor:2024iix,
    author = "Gombor, Tamas",
    title = "{Exact Overlaps for All Integrable Matrix Product States of Rational Spin Chains}",
    eprint = "2410.23282",
    archivePrefix = "arXiv",
    primaryClass = "hep-th",
    doi = "10.1103/4vy2-8cnk",
    journal = "Phys. Rev. Lett.",
    volume = "135",
    number = "15",
    pages = "150402",
    year = "2025"
}

@article{Gombor:2021hmj,
    author = "Gombor, Tam{\'a}s",
    title = "{On exact overlaps for gl(N) symmetric spin chains}",
    eprint = "2110.07960",
    archivePrefix = "arXiv",
    primaryClass = "hep-th",
    doi = "10.1016/j.nuclphysb.2022.115909",
    journal = "Nucl. Phys. B",
    volume = "983",
    pages = "115909",
    year = "2022"
}

@article{Gomis:2025gzb,
    author = "Gomis, Jaume",
    title = "{The AdS/$\mathsf{C}$-$\mathsf{P}$-${\mathsf T}$ Correspondence}",
    eprint = "2507.12467",
    archivePrefix = "arXiv",
    primaryClass = "hep-th",
    month = "7",
    year = "2025"
}

@article{Ambrosino:2026ovo,
    author = "Ambrosino, Federico and Gomis, Jaume and Kannagi, Suriyah Rajalingam",
    title = "{Monodromy defects in Chern-Simons theory and Holography}",
    eprint = "2607.06669",
    archivePrefix = "arXiv",
    primaryClass = "hep-th",
    month = "7",
    year = "2026"
}

@article{Soderberg:2017oaa,
    author = {S{\"o}derberg, Alexander},
    title = "{Anomalous Dimensions in the WF O($N$) Model with a Monodromy Line Defect}",
    eprint = "1706.02414",
    archivePrefix = "arXiv",
    primaryClass = "hep-th",
    reportNumber = "UUITP-16-17",
    doi = "10.1007/JHEP03(2018)058",
    journal = "JHEP",
    volume = "03",
    pages = "058",
    year = "2018"
}

@article{Drukker:2022txy,
    author = "Drukker, Nadav and Kong, Ziwen",
    title = "{1/3 BPS loops and defect CFTs in ABJM theory}",
    eprint = "2212.03886",
    archivePrefix = "arXiv",
    primaryClass = "hep-th",
    doi = "10.1007/JHEP06(2023)137",
    journal = "JHEP",
    volume = "06",
    pages = "137",
    year = "2023"
}

@article{Drukker:2009hy,
    author = "Drukker, Nadav and Trancanelli, Diego",
    title = "{A Supermatrix model for N=6 super Chern-Simons-matter theory}",
    eprint = "0912.3006",
    archivePrefix = "arXiv",
    primaryClass = "hep-th",
    reportNumber = "HU-EP-09-30, NSF-KITP-09-120",
    doi = "10.1007/JHEP02(2010)058",
    journal = "JHEP",
    volume = "02",
    pages = "058",
    year = "2010"
}

@article{Rey:2008bh,
    author = "Rey, Soo-Jong and Suyama, Takao and Yamaguchi, Satoshi",
    title = "{Wilson Loops in Superconformal Chern-Simons Theory and Fundamental Strings in Anti-de Sitter Supergravity Dual}",
    eprint = "0809.3786",
    archivePrefix = "arXiv",
    primaryClass = "hep-th",
    reportNumber = "SNUST-080903",
    doi = "10.1088/1126-6708/2009/03/127",
    journal = "JHEP",
    volume = "03",
    pages = "127",
    year = "2009"
}

@article{Chen:2008bp,
    author = "Chen, Bin and Wu, Jun-Bao",
    title = "{Supersymmetric Wilson Loops in N=6 Super Chern-Simons-matter theory}",
    eprint = "0809.2863",
    archivePrefix = "arXiv",
    primaryClass = "hep-th",
    reportNumber = "SISSA-60-2008-EP",
    doi = "10.1016/j.nuclphysb.2009.09.015",
    journal = "Nucl. Phys. B",
    volume = "825",
    pages = "38--51",
    year = "2010"
}

@article{Drukker:2008zx,
    author = "Drukker, Nadav and Plefka, Jan and Young, Donovan",
    title = "{Wilson loops in 3-dimensional N=6 supersymmetric Chern-Simons Theory and their string theory duals}",
    eprint = "0809.2787",
    archivePrefix = "arXiv",
    primaryClass = "hep-th",
    reportNumber = "HU-EP-08-20",
    doi = "10.1088/1126-6708/2008/11/019",
    journal = "JHEP",
    volume = "11",
    pages = "019",
    year = "2008"
}

@article{Bai:2026hun,
    author = "Bai, Nan and Yang, Hui and Shao, Mao-Zhong",
    title = "{Solving for the integrable boundary states of the ABJM spin chain from $KT$-relations}",
    eprint = "2607.04203",
    archivePrefix = "arXiv",
    primaryClass = "hep-th",
    month = "7",
    year = "2026"
}

@article{Liu:2025uiu,
    author = "Liu, Yang and Wu, Junbao",
    title = "{Chiral integrable boundary states in the SU(4) alternating spin chain}",
    eprint = "2507.03489",
    archivePrefix = "arXiv",
    primaryClass = "hep-th",
    reportNumber = "CJQS-2026-002, ICTS/PCFT-25-27",
    doi = "10.1007/s11433-025-2834-3",
    journal = "Sci. China Phys. Mech. Astron.",
    volume = "69",
    number = "3",
    pages = "231011",
    year = "2026"
}

@article{Bianchi:2021snj,
    author = "Bianchi, Lorenzo and Chalabi, Adam and Proch{\'a}zka, Vladim{\'\i}r and Robinson, Brandon and Sisti, Jacopo",
    title = "{Monodromy defects in free field theories}",
    eprint = "2104.01220",
    archivePrefix = "arXiv",
    primaryClass = "hep-th",
    reportNumber = "UUITP- 16/21",
    doi = "10.1007/JHEP08(2021)013",
    journal = "JHEP",
    volume = "08",
    pages = "013",
    year = "2021"
}

@article{Giombi:2021uae,
    author = "Giombi, Simone and Helfenberger, Elizabeth and Ji, Ziming and Khanchandani, Himanshu",
    title = "{Monodromy defects from hyperbolic space}",
    eprint = "2102.11815",
    archivePrefix = "arXiv",
    primaryClass = "hep-th",
    doi = "10.1007/JHEP02(2022)041",
    journal = "JHEP",
    volume = "02",
    pages = "041",
    year = "2022"
}

@article{Drukker:2008jm,
    author = "Drukker, Nadav and Gomis, Jaume and Young, Donovan",
    title = "{Vortex Loop Operators, M2-branes and Holography}",
    eprint = "0810.4344",
    archivePrefix = "arXiv",
    primaryClass = "hep-th",
    reportNumber = "HU-EP-08-43",
    doi = "10.1088/1126-6708/2009/03/004",
    journal = "JHEP",
    volume = "03",
    pages = "004",
    year = "2009"
}

@article{Drukker:2008wr,
    author = "Drukker, Nadav and Gomis, Jaume and Matsuura, Shunji",
    title = "{Probing N=4 SYM With Surface Operators}",
    eprint = "0805.4199",
    archivePrefix = "arXiv",
    primaryClass = "hep-th",
    reportNumber = "HU-EP-08-17",
    doi = "10.1088/1126-6708/2008/10/048",
    journal = "JHEP",
    volume = "10",
    pages = "048",
    year = "2008"
}

@article{Choi:2024ktc,
    author = "Choi, Changha and Gomis, Jaume and Izquierdo Garc{\'\i}a, Raquel",
    title = "{Surface operators and exact holography}",
    eprint = "2406.08541",
    archivePrefix = "arXiv",
    primaryClass = "hep-th",
    doi = "10.1007/JHEP12(2024)195",
    journal = "JHEP",
    volume = "12",
    pages = "195",
    year = "2024"
}

@article{Holguin:2025bfe,
    author = "Holguin, Adolfo and Kawai, Hiroki",
    title = "{Integrability and conformal blocks for surface defects in $\mathcal{N}=4$ SYM}",
    eprint = "2503.09944",
    archivePrefix = "arXiv",
    primaryClass = "hep-th",
    doi = "10.1007/JHEP11(2025)043",
    journal = "JHEP",
    volume = "11",
    pages = "043",
    year = "2025"
}

@article{Chalabi:2025nbg,
    author = "Chalabi, Adam and Kristjansen, Charlotte and Su, Chenliang",
    title = "{Integrable corners in the space of Gukov-Witten surface defects}",
    eprint = "2503.22598",
    archivePrefix = "arXiv",
    primaryClass = "hep-th",
    doi = "10.1016/j.physletb.2025.139512",
    journal = "Phys. Lett. B",
    volume = "866",
    pages = "139512",
    year = "2025"
}

@article{Gukov:2006jk,
    author = "Gukov, Sergei and Witten, Edward",
    title = "{Gauge Theory, Ramification, And The Geometric Langlands Program}",
    eprint = "hep-th/0612073",
    archivePrefix = "arXiv",
    month = "12",
    year = "2006"
}

@article{Gukov:2008sn,
    author = "Gukov, Sergei and Witten, Edward",
    title = "{Rigid Surface Operators}",
    eprint = "0804.1561",
    archivePrefix = "arXiv",
    primaryClass = "hep-th",
    doi = "10.4310/ATMP.2010.v14.n1.a3",
    journal = "Adv. Theor. Math. Phys.",
    volume = "14",
    number = "1",
    pages = "87--178",
    year = "2010"
}

@article{Kristjansen:2021abc,
    author = "Kristjansen, Charlotte and Vu, Dinh-Long and Zarembo, Konstantin",
    title = "{Integrable domain walls in ABJM theory}",
    eprint = "2112.10438",
    archivePrefix = "arXiv",
    primaryClass = "hep-th",
    reportNumber = "NORDITA 2021-151",
    doi = "10.1007/JHEP02(2022)070",
    journal = "JHEP",
    volume = "02",
    pages = "070",
    year = "2022"
}

@article{Gombor:2022aqj,
    author = "Gombor, Tamas and Kristjansen, Charlotte",
    title = "{Overlaps for matrix product states of arbitrary bond dimension in ABJM theory}",
    eprint = "2207.06866",
    archivePrefix = "arXiv",
    primaryClass = "hep-th",
    doi = "10.1016/j.physletb.2022.137428",
    journal = "Phys. Lett. B",
    volume = "834",
    pages = "137428",
    year = "2022"
}

@article{Gombor:2020kgu,
    author = "Gombor, Tamas and Bajnok, Zoltan",
    title = "{Boundary states, overlaps, nesting and bootstrapping AdS/dCFT}",
    eprint = "2004.11329",
    archivePrefix = "arXiv",
    primaryClass = "hep-th",
    doi = "10.1007/JHEP10(2020)123",
    journal = "JHEP",
    volume = "10",
    pages = "123",
    year = "2020"
}

@article{Gombor:2020auk,
    author = "Gombor, Tamas and Bajnok, Zoltan",
    title = "{Boundary state bootstrap and asymptotic overlaps in AdS/dCFT}",
    eprint = "2006.16151",
    archivePrefix = "arXiv",
    primaryClass = "hep-th",
    doi = "10.1007/JHEP03(2021)222",
    journal = "JHEP",
    volume = "03",
    pages = "222",
    year = "2021"
}

@article{Minahan:2008hf,
    author = "Minahan, J. A. and Zarembo, K.",
    title = "{The Bethe ansatz for superconformal Chern-Simons}",
    eprint = "0806.3951",
    archivePrefix = "arXiv",
    primaryClass = "hep-th",
    reportNumber = "ITEP-TH-30-08, LPTENS-08-32, UUITP-13-08",
    doi = "10.1088/1126-6708/2008/09/040",
    journal = "JHEP",
    volume = "09",
    pages = "040",
    year = "2008"
}

@article{Minahan:2009te,
    author = "Minahan, J. A. and Schulgin, W. and Zarembo, K.",
    title = "{Two loop integrability for Chern-Simons theories with N=6 supersymmetry}",
    eprint = "0901.1142",
    archivePrefix = "arXiv",
    primaryClass = "hep-th",
    reportNumber = "ITEP-TH-01-09, LPTENS-09-01, UUITP-01-09",
    doi = "10.1088/1126-6708/2009/03/057",
    journal = "JHEP",
    volume = "03",
    pages = "057",
    year = "2009"
}

@article{Komatsu:2020sup,
    author = "Komatsu, Shota and Wang, Yifan",
    title = "{Non-perturbative defect one-point functions in planar $\mathcal{N}=4$  super-Yang-Mills}",
    eprint = "2004.09514",
    archivePrefix = "arXiv",
    primaryClass = "hep-th",
    doi = "10.1016/j.nuclphysb.2020.115120",
    journal = "Nucl. Phys. B",
    volume = "958",
    pages = "115120",
    year = "2020"
}

@article{deLeeuw:2015hxa,
    author = "de Leeuw, Marius and Kristjansen, Charlotte and Zarembo, Konstantin",
    title = "{One-point Functions in Defect CFT and Integrability}",
    eprint = "1506.06958",
    archivePrefix = "arXiv",
    primaryClass = "hep-th",
    reportNumber = "NORDITA-2015-72, UUITP-12-15",
    doi = "10.1007/JHEP08(2015)098",
    journal = "JHEP",
    volume = "08",
    pages = "098",
    year = "2015"
}

@article{Buhl-Mortensen:2015gfd,
    author = "Buhl-Mortensen, Isak and de Leeuw, Marius and Kristjansen, Charlotte and Zarembo, Konstantin",
    title = "{One-point Functions in AdS/dCFT from Matrix Product States}",
    eprint = "1512.02532",
    archivePrefix = "arXiv",
    primaryClass = "hep-th",
    reportNumber = "NORDITA-2015-132, UUITP-26-15",
    doi = "10.1007/JHEP02(2016)052",
    journal = "JHEP",
    volume = "02",
    pages = "052",
    year = "2016"
}

@article{Buhl-Mortensen:2016pxs,
    author = "Buhl-Mortensen, Isak and de Leeuw, Marius and Ipsen, Asger C. and Kristjansen, Charlotte and Wilhelm, Matthias",
    title = "{One-loop one-point functions in gauge-gravity dualities with defects}",
    eprint = "1606.01886",
    archivePrefix = "arXiv",
    primaryClass = "hep-th",
    doi = "10.1103/PhysRevLett.117.231603",
    journal = "Phys. Rev. Lett.",
    volume = "117",
    number = "23",
    pages = "231603",
    year = "2016"
}

@article{Buhl-Mortensen:2016jqo,
    author = "Buhl-Mortensen, Isak and de Leeuw, Marius and Ipsen, Asger C. and Kristjansen, Charlotte and Wilhelm, Matthias",
    title = "{A Quantum Check of AdS/dCFT}",
    eprint = "1611.04603",
    archivePrefix = "arXiv",
    primaryClass = "hep-th",
    doi = "10.1007/JHEP01(2017)098",
    journal = "JHEP",
    volume = "01",
    pages = "098",
    year = "2017"
}

@article{Gombor:2024api,
    author = "Gombor, Tamas and Bajnok, Zolt{\'a}n",
    title = "{Dual overlaps and finite coupling {\textquoteright}t Hooft loops}",
    eprint = "2408.14901",
    archivePrefix = "arXiv",
    primaryClass = "hep-th",
    doi = "10.1007/JHEP12(2024)034",
    journal = "JHEP",
    volume = "12",
    pages = "034",
    year = "2024"
}

@article{Bandres_2008,
   title={Studies of the ABJM theory in a formulation with manifest SU(4) R-symmetry},
   volume={2008},
   ISSN={1029-8479},
   url={http://dx.doi.org/10.1088/1126-6708/2008/09/027},
   DOI={10.1088/1126-6708/2008/09/027},
   number={09},
   journal={Journal of High Energy Physics},
   publisher={Springer Science and Business Media LLC},
   author={Bandres, Miguel A and Lipstein, Arthur E and Schwarz, John H},
   year={2008},
   month=sep, pages={027–027} }

@article{Yang:2021hrl,
    author = "Yang, Peihe and Jiang, Yunfeng and Komatsu, Shota and Wu, Jun-Bao",
    title = "{Three-point functions in ABJM and Bethe Ansatz}",
    eprint = "2103.15840",
    archivePrefix = "arXiv",
    primaryClass = "hep-th",
    reportNumber = "CERN-TH-2021-042, USTC-ICTS/PCFT-21-14, CJQS-2022-001",
    doi = "10.1007/JHEP01(2022)002",
    journal = "JHEP",
    volume = "01",
    pages = "002",
    year = "2022"
}

@article{Bai:2024qtg,
    author = "Bai, Nan and Shao, Mao-Zhong",
    title = "{Integrable matrix product states of ABJM theory from projecting method}",
    eprint = "2411.09282",
    archivePrefix = "arXiv",
    primaryClass = "hep-th",
    doi = "10.1142/S0217732325500683",
    journal = "Mod. Phys. Lett. A",
    volume = "40",
    number = "19n20",
    pages = "2550068",
    year = "2025"
}

\end{document}